\documentclass[a4paper,11pt]{article}
\pdfoutput=1 
\usepackage{jinstpub} 
\usepackage{natbib} 

\title{\boldmath Characterization of the Multi-Blade 10B-based detector at the CRISP reflectometer at ISIS for neutron reflectometry at ESS}

\author[a,1]{F. Piscitelli,\note{Corresponding author.}}
\author[b,a]{G. Mauri,}
\author[c,a]{F. Messi,}
\author[a]{M. Anastasopoulos,}
\author[a]{T. Arnold,}
\author[d]{A. Glavic,}
\author[a,e]{C. H\"{o}glund,}
\author[c]{T. Ilves,}
\author[a]{I. Lopez Higuera,}
\author[f]{P. Pazmandi,}
\author[g]{D. Raspino,}
\author[a]{L. Robinson,}
\author[a,h]{S. Schmidt,}
\author[a]{P. Svensson,}
\author[f]{D. Varga,}
\author[a,i]{R. Hall-Wilton}

\affiliation[a]{European Spallation Source ERIC (ESS), P.O. Box 176, SE-22100 Lund, Sweden.}
\affiliation[b]{Department of Physics, University of Perugia, Piazza Universit\'a 1, 06123 Perugia, Italy.}
\affiliation[c]{Division of Nuclear Physics, Lund University, P.O. Box 118, SE-22100 Lund, Sweden.}
\affiliation[d]{Paul Scherrer Institute, WHGA/530, 5232 Villigen - PSI, Switzerland.}
\affiliation[e]{Department of Physics, Chemistry and Biology, Link\"{o}ping University, SE-58183 Link\"{o}ping, Sweden.}
\affiliation[f]{Wigner Research Centre for Physics, Konkoly Thege Mikl\'os \'ut 29-33, H-1121 Budapest, Hungary.}
\affiliation[g]{ISIS Neutron and Muon Source, Harwell Oxford, Didcot OX11 0QX, United Kingdom.}
\affiliation[h]{IHI Ionbond AG, Industriestrasse 211, 4600 Olten, Switzerland.}
\affiliation[i]{Mid-Sweden University, SE-85170 Sundsvall, Sweden.}

\emailAdd{francesco.piscitelli@esss.se}

\abstract{The Multi-Blade is a Boron-10-based gaseous thermal neutron detector developed to face the challenge arising in neutron reflectometry at neutron sources. Neutron reflectometers are challenging instruments in terms of instantaneous counting rate and spatial resolution. This detector has been designed according to the requirements given by the reflectometers at the European Spallation Source (ESS) in Sweden. The Multi-Blade has been installed and tested on the CRISP reflectometer at the ISIS neutron and muon source in UK. The results on the detailed detector characterization are discussed in this manuscript.}


\arxivnumber{xxx.xxx} 

\begin{document}
\maketitle
\flushbottom

\section{Introduction}
The Multi-Blade~\cite{MIO_MB2017,MIO_MB2014,MIO_MBproc,MIO_MyThesis} is a gaseous thermal neutron detector based on $\mathrm{^{10}B}$ solid converters for neutron reflectometers. The design of the Multi-Blade detector has been driven by the requirements set by the two reflectometers foreseen at the European Spallation Source (ESS~\cite{ESS,ESS_TDR}) in Sweden: FREIA~\cite{INSTR_FREIA,INSTR_FREIA2} (horizontal reflectometer) and ESTIA~\cite{INSTR_ESTIA,INSTR_ESTIA1,INSTR_ESTIA2} (vertical reflectometer). 
\\ Neutron reflectometers are a challenging class of instruments in terms of detector requirements. The need for better performance in addition to the scarcity of $\mathrm{^3He}$~\cite{HE3S_kouzes,HE3S_kouzes2} are both driving the developments of new detector technology for neutron scattering science in general. The key detector requirements for neutron reflectometers are the counting rate capability and the spatial resolution. These are essential features for the detectors at the ESS reflectometers, whereby the expected instantaneous local flux is about $10^{5}/s/mm^2$~\cite{ESS_TDR,DET_rates,HE3S_kirstein,HE3S_cooper} and a sub-mm spatial resolution (Full-Width-Half-Maximum, FWHM) is required.
\\ The state-of-the-art detector technology is reaching fundamental limits. The spatial resolution is limited to approximately $2 \times 8\,mm^2$, with a maximum counting rate capability of $40\,kHz$ integrated over the whole beam intensity on the detector~\cite{INSTR_FIGARO}. The latter corresponds to a few hundreds $Hz/mm^2$. Note that at the incident rates available at existing sources (pulsed and reactors) detectors already saturate. 
\\ At current facilities the time resolution for kinetic studies is limited by the available flux and by the detector performance. In order to open the possibility of sub-second kinetic studies, a new instrument layout, which exploits a higher neutron flux, has been presented for reactor sources~\cite{R_Cubitt1,R_Cubitt2} and it requires high spatial resolution detectors. Therefore, a more performing detector technology is needed, not only for the ESS reflectometers, but also at existing reflectometers at current facilities. The Multi-Blade detector is designed to fulfil these challenging requirements. 
\\ Table~\ref{tab1} summarizes the detector requirements for the two ESS reflectometers. 
\begin{table}[htbp]
\centering
\caption{\label{tab1} \footnotesize Detector requirements for neutron reflectometers at ESS.}
\smallskip
\begin{tabular}{|l|l|l|}
\hline
\hline
  & FREIA~\cite{INSTR_FREIA,INSTR_FREIA2} & ESTIA~\cite{INSTR_ESTIA,INSTR_ESTIA1}  \\        
\hline
\hline
wavelength range (\AA) &  2.5 - 12  & 4 - 10 \\
\hline
minimum detection efficiency  &  >40\% at 2.5\AA  & >45\% at 4\AA \\
\hline
sample-detector distance (m) & 3 & 4 \\
\hline
instantaneous local rate & & \\
on detector ($kHz/mm^2$) & $100$  & $100$ \\
\hline
sensitive area: x-direction (horizontal) ($mm$)  & 300 & 500 \\
sensitive area: y-direction (vertical) ($mm$)  & 300 & 250 \\
\hline
spatial resolution (FWHM) x ($mm$)  & 2.5 & 0.5 \\
spatial resolution (FWHM) y ($mm$)  & 0.5 & 4 \\
\hline
uniformity ($\%$)  & 5 & 5 \\
\hline
desired max window scattering  & $10^{-4}$ & $10^{-4}$ \\
 \hline
$\gamma$-ray sensitivity & $<10^{-6}$ & $<10^{-6}$ \\
 \hline
 \hline
\end{tabular}
\end{table}
\\ The Multi-Blade detector has been previously characterized and several demonstrators have been built. A detailed description of the detector can be found in~\cite{MIO_MB2017,MIO_MB2014,MIO_MBproc,MIO_MyThesis} along with the results of the characterization carried out at the Budapest Neutron Centre (BNC)~\cite{FAC_BNC} and at the Source Testing Facility (STF)~\cite{SF2,SF1} at the Lund University in Sweden. It has been shown that this detector technology represents a valid alternative to the state-of-the-art technology for neutron reflectometry instruments that use cold neutrons (2.5-30\AA~\cite{INSTR_FIGARO,INSTR_D17}). Most of the detector requirements have already been fulfilled by the Multi-Blade technology. A spatial resolution of $\approx 0.6\,mm$ has been measured together with a detection efficiency ($\approx 44\%$ at the shortest wavelength 2.5\AA). The counting rate capability of this detector has been measured up to $\approx 17\,kHz/ch$ limited by the available neutron flux at the source. The gamma-ray sensitivity~\cite{MIO_MB2017,MG_gamma} below $10^{-7}$ has been measured with the Multi-Blade detector as well as the sensitivity to fast neutrons ($1-10\,MeV$) of approximately $10^{-5}$~\cite{MIO_fastn}.
\\ A Multi-Blade detector has been installed and tested on the CRISP~\cite{CRISP1} reflectometer at ISIS (Science \& Technology Facilities Council in UK~\cite{ISIS}). The aim of this test was to make a detector technology demonstration on a neutron reflectometry instrument. These tests are crucial to validate the Multi-Blade technology to be installed at the ESS reflectometers. The reflectivity of several reference samples have also been measured with the Multi-blade detector at the instrument but this is not the subject of this manuscript which is focused on the technical aspects of the detector and its characterization. The scientific measurements carried out with the Multi-Blade on CRISP will be part of a different paper.
\section{Description of the setup}\label{MBtes}
\subsection{The Multi-Blade detector}
\begin{figure}[htbp]
\centering
\includegraphics[width=.8\textwidth,keepaspectratio]{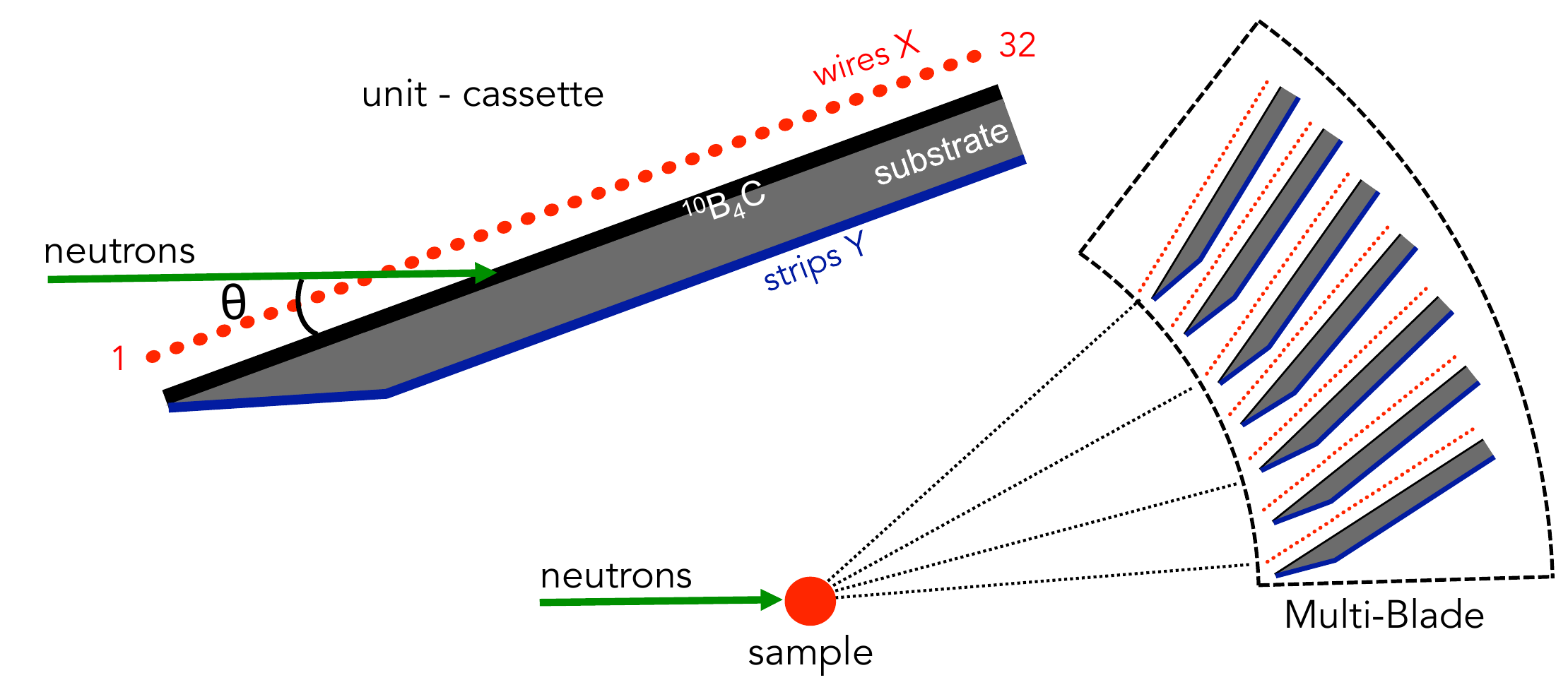}
\caption{\label{fig99} \footnotesize Schematic view of the cross-section of the Multi-Blade detector made up of identical units (cassettes) arranged adjacent to each other. Please note that the scale is exaggerated for ease of viewing. Each cassette holds a $\mathrm{^{10}B_4C}$-layer; the readout is performed through a plane of wires and a plane of strips. Picture from~\cite{MIO_MB2017}.}
\end{figure}

\begin{figure}[htbp]
\centering
\includegraphics[width=.8\textwidth,keepaspectratio]{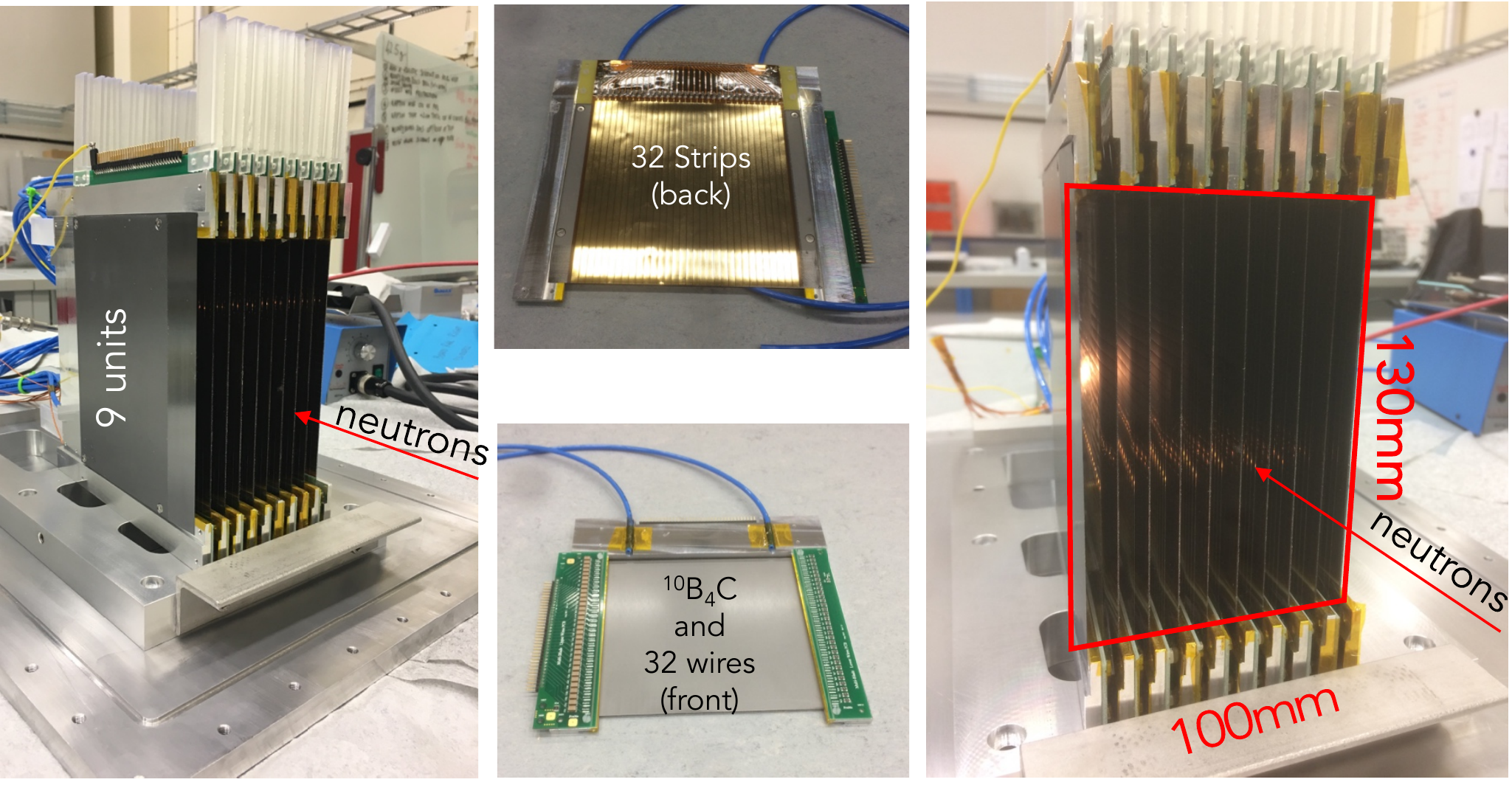}
\caption{\label{fig1} \footnotesize A picture of the Multi-Blade detector with Ti-blades made up of 9 units (cassettes), left and right. A picture of the two sides of a cassettes holding the wires, the blade with the $\mathrm{^{10}B_4C}$ layer and the strips (centre).}
\end{figure}

The Multi-Blade is a stack of Multi Wire Proportional Chambers (MWPC) operated at atmospheric pressure with a continuous gas flow ($\mathrm{Ar/CO_2}$ 80/20 mixture). A sketch of the Multi-Blade detector is shown in Figure~\ref{fig99}. The Multi-Blade is made up of identical units, the so-called `cassettes'. Each cassette holds a `blade' (a flat substrate coated with $\mathrm{^{10}B_4C}$~\cite{B4C_carina,B4C_carina3,B4C_Schmidt}) and a two-dimensional readout system, which consists of a plane of wires and a plane of strips. The readout of a single converter is performed by the facing anode wire plane which mechanically belongs to the cassette and the strip plane that belongs to the adjacent cassette. Each $\mathrm{^{10}B_4C}$-converter (blade) is inclined at grazing angle ($\beta = 5$ degrees) with respect to the incoming neutron beam. The inclined geometry has two advantages: the neutron flux is shared among more wires with respect to the normal incidence (the counting rate capability is correspondingly increased) and the spatial resolution is similarly improved. Moreover, the use of the $\mathrm{^{10}B_4C}$ conversion layer at an angle also increases the detection efficiency, which is otherwise limited to a few percent at thermal energies for a single converter~\cite{MIO_analyt}. As stated in~\cite{MIO_MB2017}, the cassettes are arranged over a circle around the sample and they have some overlap; i.e. each blade makes a small shadow over the adjacent one in order to avoid dead areas.
\\Figure~\ref{fig1} shows a picture of the Multi-Blade detector. A stack of 9 units (cassette) and a picture of the front and back of a single cassette, holding the wires and strips. 
\\ The Multi-Blade described in~\cite{MIO_MB2017} exploited $2\,mm$ thick Al-blades coated with the neutron converter layer ($\mathrm{^{10}B_4C}$). For deposition on a single side of the substrate (blade), a deformation of the substrates was observed. Note that the planarity of the substrate is crucial for the uniformity of the electric field of the MWPC and for defining the position of the neutron detection. In the present detector these blades have been replaced with Titanium (Ti) or Stainless Steel (SS) of thickness of $2\,mm$, and they both show a very good response in terms of mechanical stress, i.e. planarity is not an issue even if the $\mathrm{^{10}B_4C}$ layer is deposited on a single side. The coating thickness on these blades is $4.4\,\mu m$ but it has been shown~\cite{MIO_MB2017} that a $7.5\,\mu m$-layer is needed to absorb $\approx 99\%$ of the neutrons at the shortest wavelength 2.5\AA\, of the two ESS reflectometers (any neutron of longer wavelength is absorbed with higher probability). A neutron that crosses the layer and reaches the substrate, can be scattered causing  unwanted spurious events within the detector. The detection efficiency is saturated above $3\,\mu m$ and any extra film thickness will only help to absorb neutrons~\cite{MIO_MB2017}. 
\\ The cassettes are mounted on an array plate which gives the alignment and matches the circle of $4\,m$ radius which is needed for ESTIA; this distance is measured from the centre of the sample to the first wire of each cassette (wire no. 1 in Figure~\ref{fig99}). Moreover, the cassettes are not parallel as described in~\cite{MIO_MB2017} but  the relative angle between two adjacent cassettes is $0.14$ degrees. This arrangement requires the positioning of each cassette on the array plate with a precision of approximately $0.15\,mm$.  
\\ The vessel of the present Multi-Blade detector has a $1\,mm$-thick aluminium entrance window. No neutron shielding was foreseen for such a vessel, but a borated rubber cover was improvised at the instrument during the tests. 
\subsection{Front-end and readout electronics}
\begin{figure}[htbp]
\centering
\includegraphics[width=.7\textwidth,keepaspectratio]{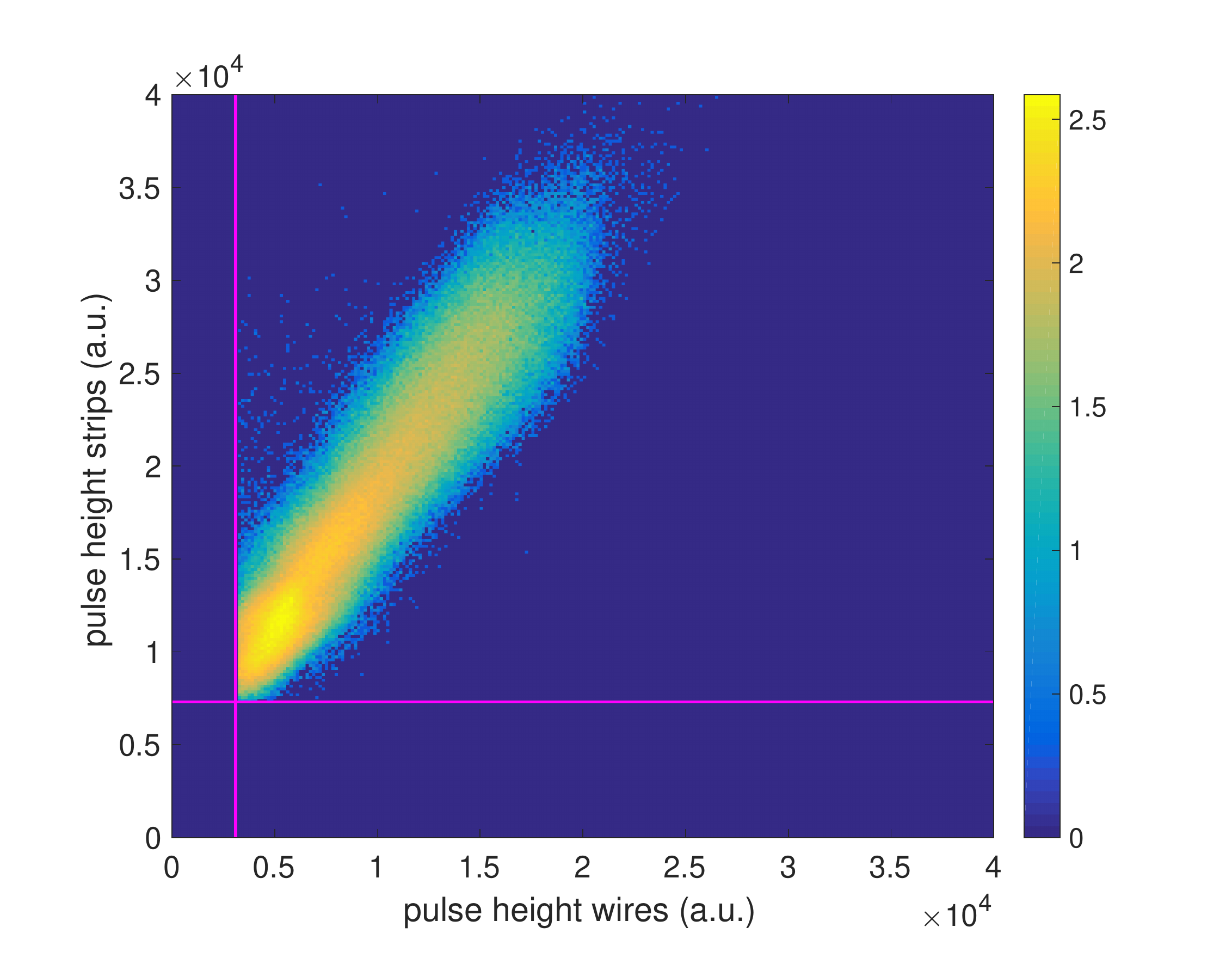}
\caption{\label{fig3334} \footnotesize Wire and strip pulse height correlation. The color scale is counts in logarithmic units. The two pink lines represent the hardware threshold applied to the readout channels to reject the electronic noise.}
\end{figure}

\begin{figure}[htbp]
\centering
\includegraphics[width=.98\textwidth,keepaspectratio]{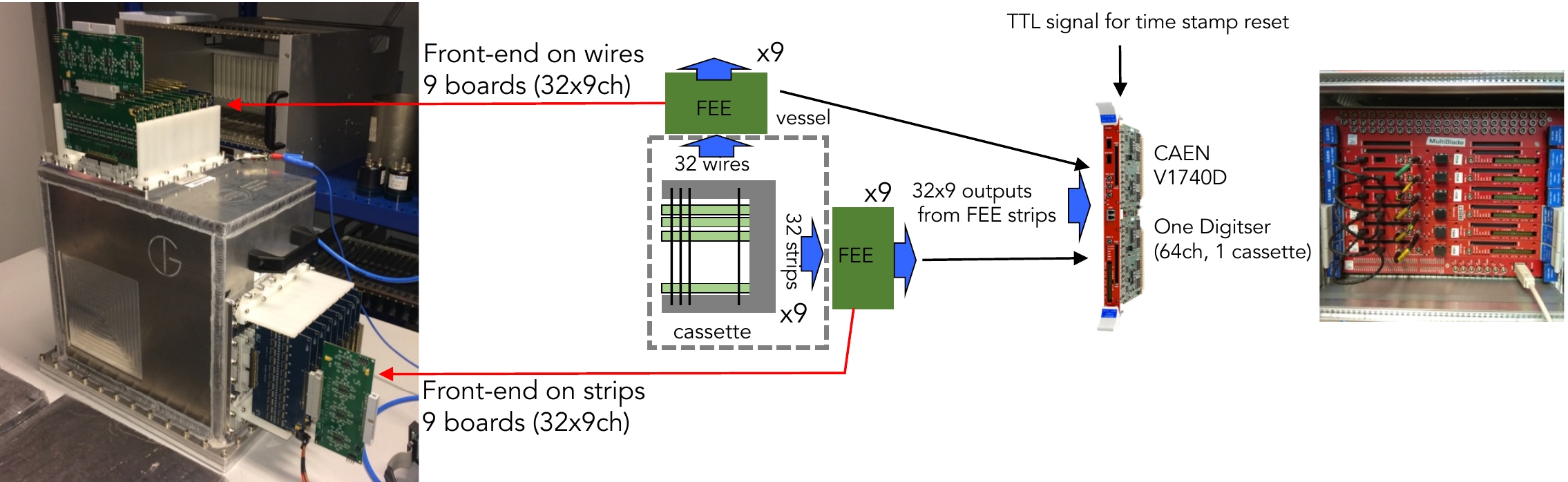}
\caption{\label{fig3} \footnotesize A picture and a sketch of the readout electronics scheme of the Multi-Blade detector. Individual readout boards (32 channels) based on FET are connected to each pane of wires and strips; each board is connected to a CAEN V1740D digitizer which allows the reset of the time-stamp.}
\end{figure}

\begin{figure}[htbp]
\centering
\includegraphics[width=.7\textwidth,keepaspectratio]{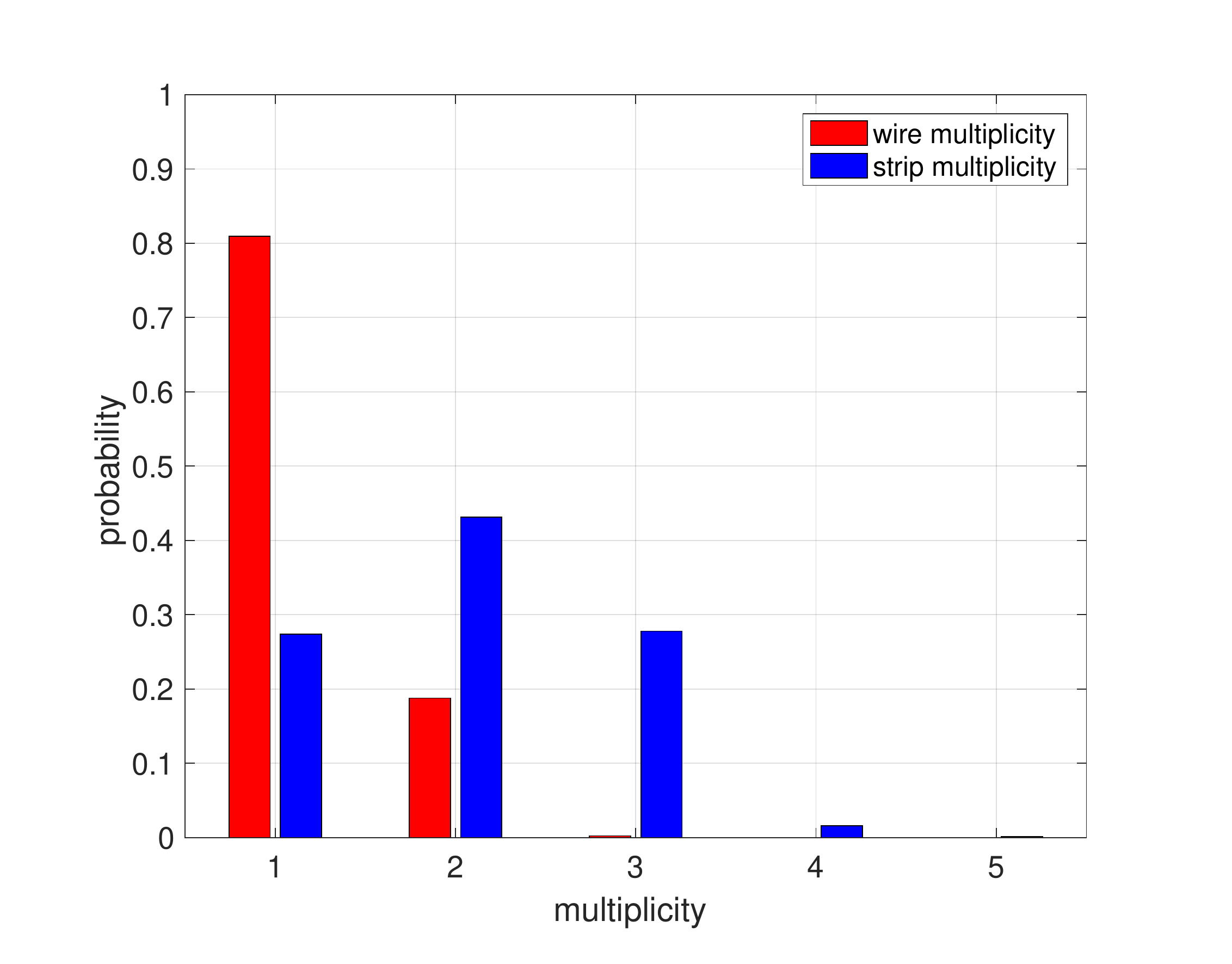}
\caption{\label{fig47} \footnotesize Probability of each multiplicity of wires and strips in a cluster for detected thermal neutron. The multiplicity is normalized to 1 and defined as the number of wires or strips respectively with a detected signal above the software threshold.}
\end{figure}

With respect to the previous demonstrator described in~\cite{MIO_MB2017}, the new detector does not employ any charge division readout~\cite{DET_chargediv} and each channel (64 per cassette, 32 wires and 32 strips) is read out individually. Each channel is connected to a FET-based charge pre-amplifier and shaper with an approximate gain of $10\,V/pC$ and shaping time of $\approx 2\,\mu s$. There is a front-end electronics (FEE) board (32 channels) connected to each plane of wires and each plane of strips. Figure~\ref{fig3} shows a sketch of the readout system. The present detector consists of 9 units (576 channels in total). The individual readout is the sole scheme that can lead to a high counting rate applications for two main reasons: global dead time is reduced since each channel is independent from the others, the amount of charge needed to perform the individual readout is generally lower than that of charge division to achieve the same signal-to-noise. Moreover, less charge means smaller space charge effects~\cite{RC_Ivaniouchenkov,RC_mathieson1,RC_mathieson2} which affects the gas gain variation of the detector at high rates. Note that, at high rate operation, the individual readout (as opposed to charge division) is mandatory to disentangle hits occurring nearly at the same time (that is, unresolved due to the finite time resolution of the detector). The measured amplitudes on the wires and on the strips are strongly correlated (since they are induced by the same avalanche), therefore with sufficient dynamic range, the ambiguity might be resolved by requiring matching amplitudes. Figure~\ref{fig3334} shows the correlation between the pulse height of strips and that of wires. The ratio between the amplitudes on wires and strips depends on the amount of charge collected at the electrode and the gain of the front-end channel. In Figure~\ref{fig3334} the amplitudes on wires are cut below approximately $3\times10^3$ a.u. (vertical pink line in the plot) and on strips below approximately $7.5\times10^3$ a.u. (horizontal pink line in the plot) which correspond to $5\,mV$ and $15\,mV$ respectively (or approximately $0.2 - 0.3\, fC$). These values are the hardware thresholds set to reject the electronic noise. 
\\ Each 32-channel board is connected to a CAEN V1740D digitizer ($12\,bit$, $62.5\,MS/s$)~\cite{EL_CAEN}. There are 6 digitizers in total and each can readout 64 channels , i.e. one cassette. Thus, out of the 9 cassettes only 6 could be used simultaneously in the tests. The 6 digitizers can be synchronized to the same clock source and a TTL logic signal can be sent to one of them and propagated to reset the time-stamp which is associated to an event. This feature is needed to perform any type of Time-of-Flight (ToF) measurement. The system is asynchronous and each time any channel is above the set hardware threshold the digitizer calculates the area of the trace in a given gate and records this event to file with the relative time-stamp. Since the signals are shaped, any value among amplitude, area of the pulse or time-over-threshold (ToT) give the same information: a value related to the energy released on a wire or a strip. 
\\ The raw data, containing the channel number and its time-stamp, is reduced to clusters of groups of channels. A software threshold can be applied to each channel in order to reject background events. The software thresholds used in these measurements corresponds to approximately $100\,keV$~\cite{MIO_MB2017,MIO_fastn}. We define a cluster a group of events in the file with the same time-stamp within a time window of $6\, \mu s$. This window is defined from the first active wire of a cluster since the strip signals arrive at the same time or later. The information of a cluster, or a hit, is given by a triplet such as (X,Y,ToF). 
\\ As already shown in~\cite{MIO_MB2017}, each neutron event could have a multiplicity more than one; each time a neutron is converted about $75\, \%$ of the times a single wire is involved in the detection process, about $25\, \%$ two wires are firing; the probability to get three or more wires involved in a detection process is below $1\, \%$. About $25\, \%$ of the times only one or three strips are involved in a detection process whereas $50\, \%$ of the times two strips are firing at the same time. The probability to get four or more strips firing is below $1\, \%$. The most probable cluster in the Multi-Blade geometry is then a wire and two strips firing at the same time. The multiplicity can be used to discriminate against background events; for instance gamma-rays interactions and fast neutron interactions have in general higher multiplicity due to the longer range in gas of electrons and protons~\cite{MIO_fastn}. 
\\ Note that the multiplicity depends on the applied thresholds and the gas gain at which the detector is operated. Hence a systematic variation of
these values are expected, without qualitative change of the confirmed behaviour. Figure~\ref{fig47} shows the probability of each multiplicity of wires and strips in a cluster obtained in these tests. 
\\ The fact that wires and strips have in general different multiplicity~\cite{DET_Stability_Bloom} is due to the combination of two phenomena: the actual extension in space of the neutron capture fragment tracks ($\alpha$ and Li particles) which in our case is comparable to the wire pitch and strip width ($4\,mm$); and the physical process of induction of the signal from the avalanche at the various electrodes. The first is responsible of the spread of the charge among more than one wire; if the ionization track of one of the neutron capture fragments is shared between two wires there will be two avalanche processes, one at each wire. Note that on all the other wires, which are not taking part to the detection process, will be induced a bipolar signal with amplitude proportional to their distance to the firing wire(s). The second phenomenon is due to the physical induction mechanism on the electrodes: even if only one wire receives the whole charge, this does not happen for the facing strips that share the charge. It is demonstrated less than $50\,\%$ is the fraction of signal measured on an infinitely wide strip~\cite{DET_Stability_Bloom}. Moreover, for narrower strips this fraction is still quite large, e.g. in our case of strip width of $4\,mm$ and comparable wire-strip distance, $\approx 28\,\%$ is collected when the avalanche is at the centre of a strip and $\approx 14\,\%$ if it is over the centre of the neighbouring strip. Note that the best position resolution is achieved when the strip width is matching the wire-strip distance~\cite{DET_gatti}, as in the Multi-Blade. When the coordinate of the avalanche is exactly in between two strips, it induces two equal signals on them. If the avalanche is produced in correspondence with the centre of one strip, it induces a large signal on it and two smaller signals on the two adjacent strips. Typically one records two signals in the first case and three in the second. Multiplicity of 1 on strips can be obtained when three signals are induced, but the two side strips do not receive enough charge to generate a signal above the electronic noise. It appears clear then how the multiplicity 1 and 3 on strips, in the Multi-Blade geometry, add up to $\approx50\, \%$ as it is for the multiplicity 2. 
\begin{figure}[htbp]
\centering
\includegraphics[width=.45\textwidth,keepaspectratio]{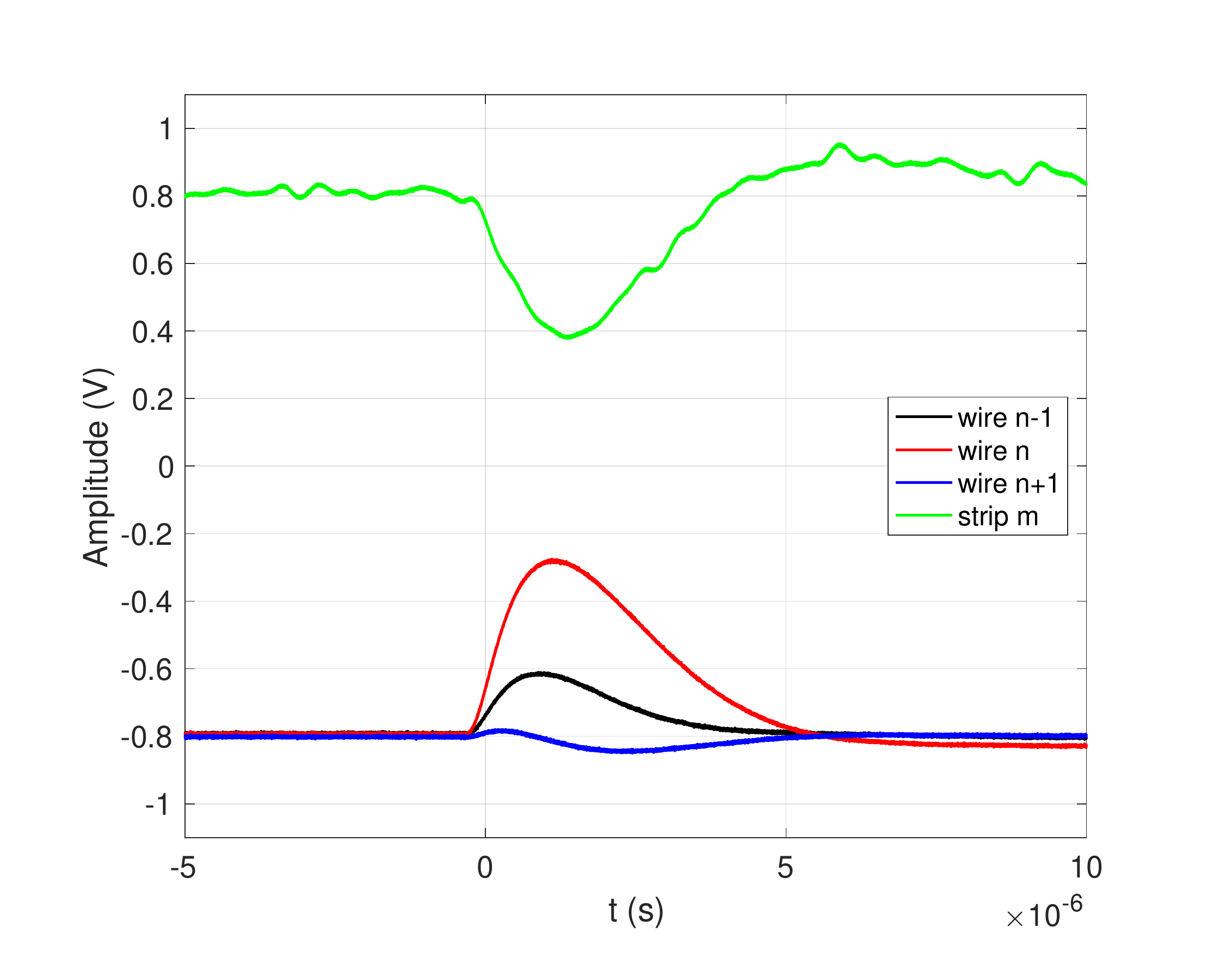}
\includegraphics[width=.45\textwidth,keepaspectratio]{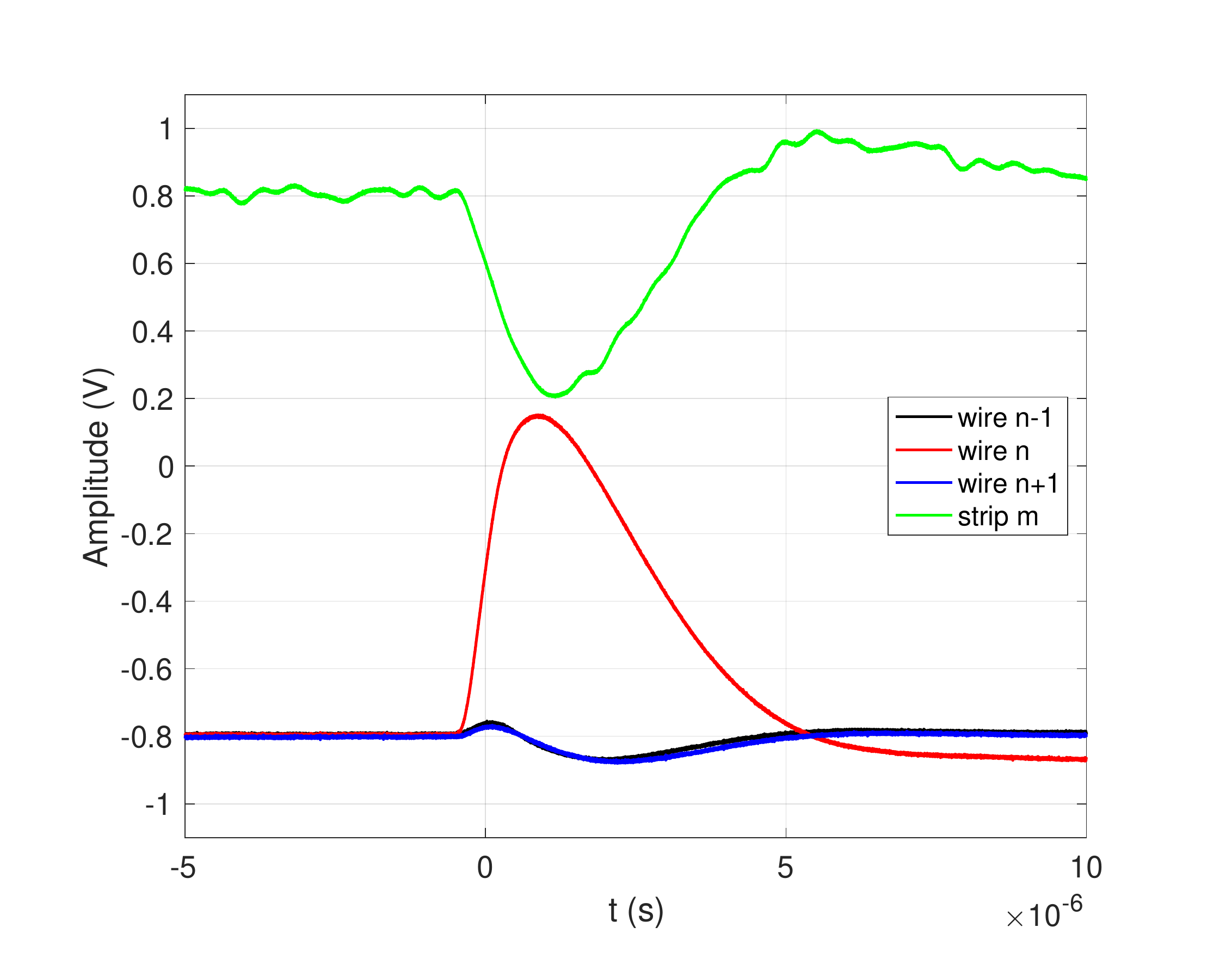}\\
\includegraphics[width=.45\textwidth,keepaspectratio]{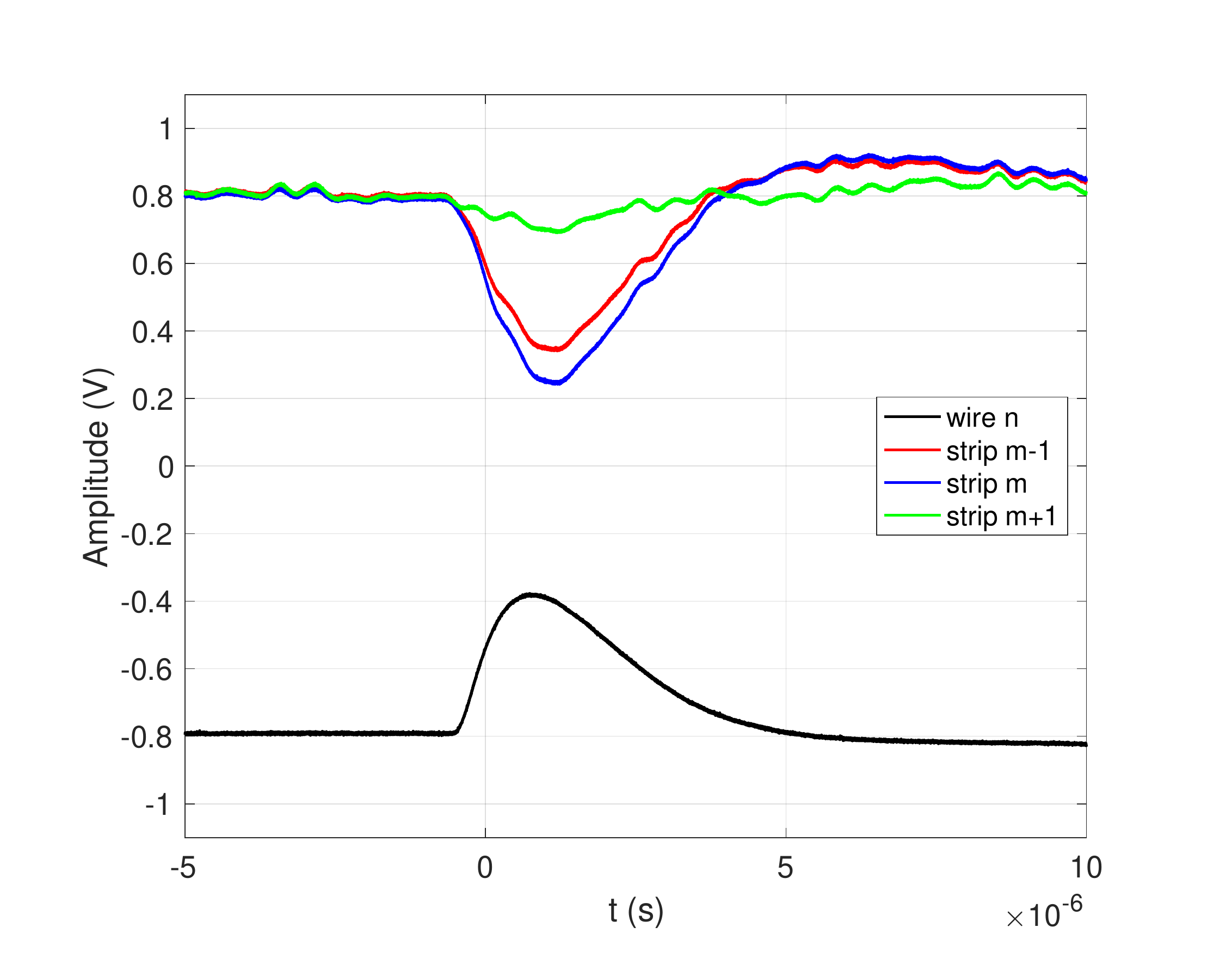}
\includegraphics[width=.45\textwidth,keepaspectratio]{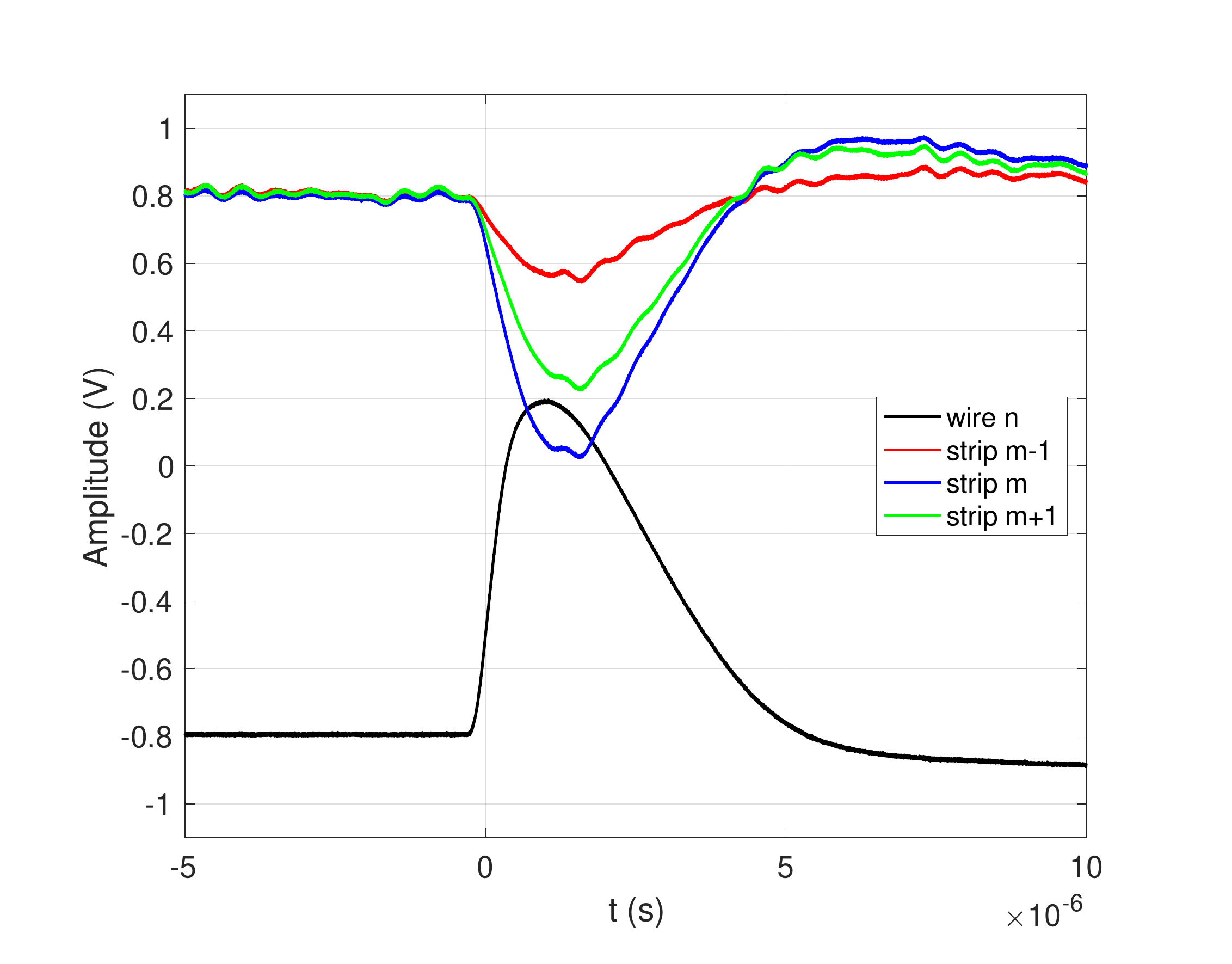}
\caption{\label{fig3b} \footnotesize Typical examples of the signals from wires and strips when firing due to a neutron interaction. Three adjacent wires and one strip with an event having multiplicity 2 (top-left) and 1 (top-right) on wires. Three adjacent strips and one wire with an event having multiplicity 2 (bottom-left) and 3 (bottom-right) on strips. The signals are all inverted due to the use of inverting amplifiers. Strip and wire signals have been shifted arbitrarily on the y-axis for viewing purposes.}
\end{figure} 
\\ Figure~\ref{fig3b} shows some examples of signals from the FEE boards from adjacent wires and strips when a neutron is detected. The two top plots show the signals of a strip and three adjacent wires when two wires are firing (wire multiplicity 2) and when only one wire is firing (wire multiplicity 1). The two bottom plots show the signals of a wire and three adjacent strips when two strips are firing (strip multiplicity 2) and when three strips are firing (strip multiplicity 3). 

The clusters, i.e. triplets, define a three-dimensional space containing the information where the neutron was detected with associated ToF, i.e. its wavelength. 
\\ Referring to Figure~\ref{fig4}, the spatial coordinates, X and Y, of a triplet, reflect the physical channels in the detector: 32 strips and 32 wires respectively. The spatial coordinates, X and Y, represent the projection over the detector entrance window ( i.e. the projection of the blades toward the neutron incoming direction). Note that the ToF of each triplet is the time of arrival of that neutron to the specific wire. If the ToF has to be encoded in neutron wavelength ($\lambda$), the physical position of each wire in depth (Z) must be taken into account. The flight path must be corrected with the distance ($Z_i$) of the wire $i-th$ of each cassette according this formula:  
\begin{equation}
\label{equadep}
D_i = D_0+Z_i = D_0+(Y_i - 1)\cdot(p\cdot\cos(\beta))
\end{equation}
where $D_0$ depends on the instrument geometry and in our case is the distance from moderator to the first wire (front wire) of the Multi-Blade corresponding to $Y_1=1$, $p=4\,mm$ is the wire pitch and $\beta=5^o$ is the inclination of each blade with respect to the sample position.  
\subsection{The setup at the CRISP reflectometer}\label{rifle}
\begin{figure}[htbp]
\centering
\includegraphics[width=.8\textwidth,keepaspectratio]{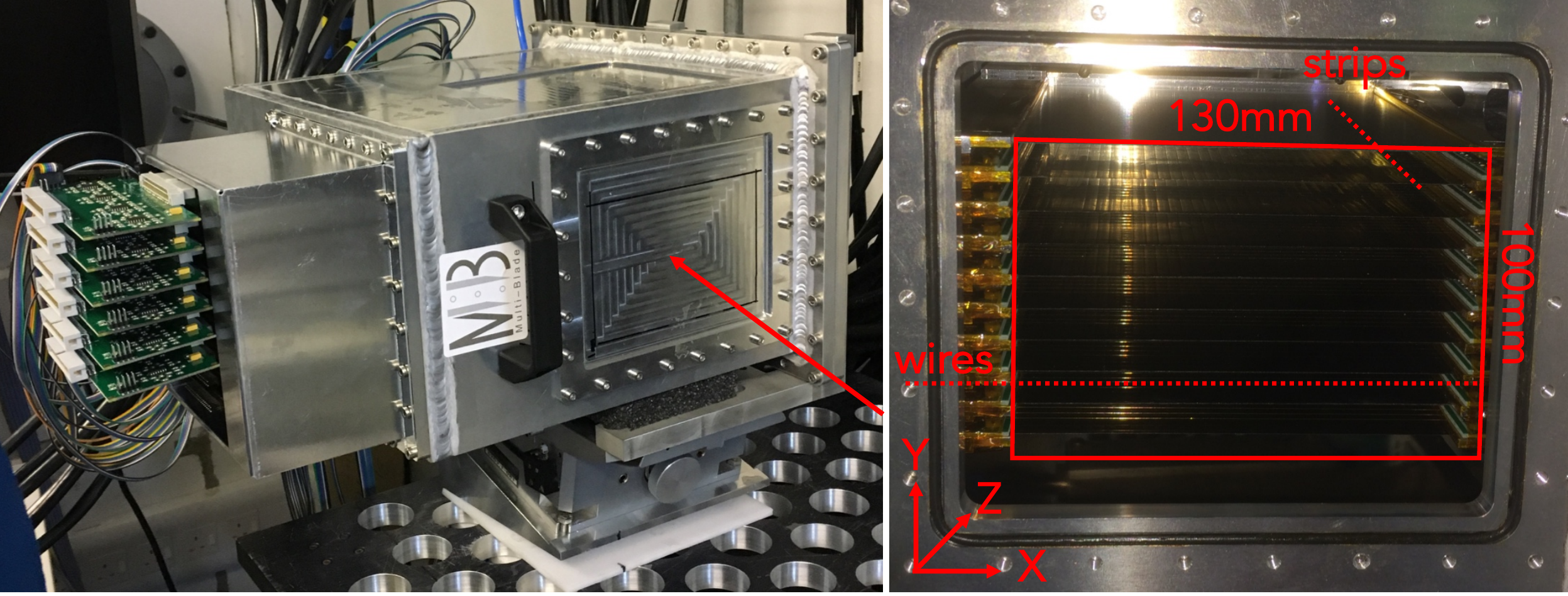}
\caption{\label{fig4} \footnotesize The Multi-Blade installed on the table of CRISP on a goniometer with the FEE boards (left). A view of the cassettes behind the vessel window to show the orientation of the wires and strips in space (right).}
\end{figure}

CRISP is an horizontal neutron reflectometer on TS1 at ISIS that uses a broad band neutron Time-of-Flight (ToF) method for determining the neutron wavelength ($\lambda$), and hence the wave vector transfer ($q$), at fixed angles ($\theta$), the reflected beam angle at the sample surface. A detailed description of the CRISP reflectometer can be found in~\cite{CRISP1}.
\\ The instrument views an hydrogen moderator giving an effective wavelength range of $0.5-6.5$\AA\, at the source frequency of $50\,Hz$. The wavelength band extends up to 13\,\AA\, if operated at $25\,Hz$. A frame overlap mirror suppress the wavelength above 13\,\AA. The distance from the moderator to the sample is $10.25\,m$ and the sample to detector distance is $1.87\,m$. The detector is a single $\mathrm{^3He}$ tube filled with $3.5\,bar$ $\mathrm{^3He}$. The Multi-Blade detector was installed at a distance of $2.3\,m$. 
Figure~\ref{fig4} shows the Multi-Blade installed on CRISP and the orientation of the blades behind the vessel entrance window. 
\\ The Multi-Blade detector was electrically insulated from the moving stage underneath in order to decouple the detector and the motion unit grounds.
\section{Results}
\subsection{Threshold choice and scattering from substrate}\label{scatt}
\begin{figure}[htbp]
\centering
\includegraphics[width=.49\textwidth,keepaspectratio]{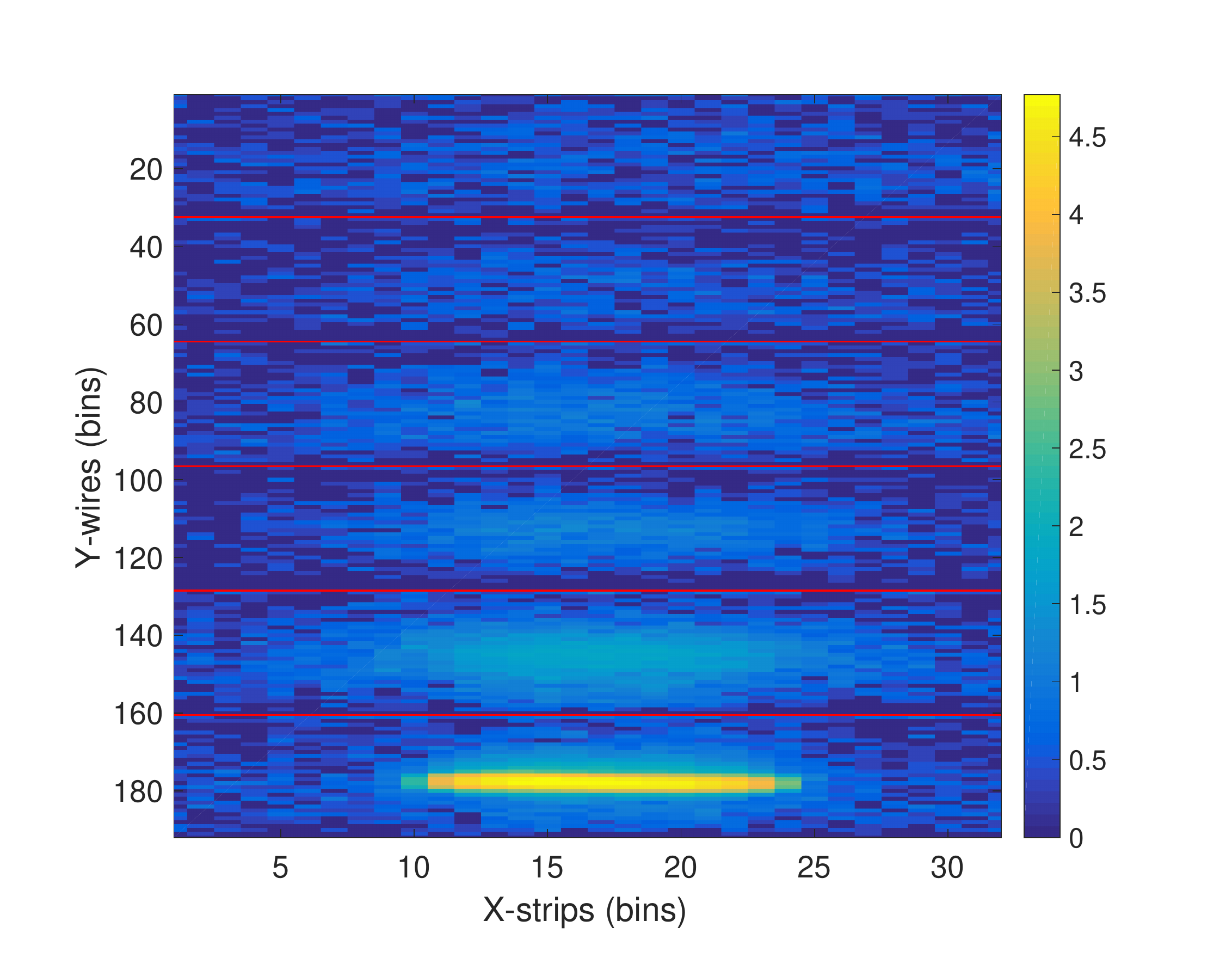}
\includegraphics[width=.49\textwidth,keepaspectratio]{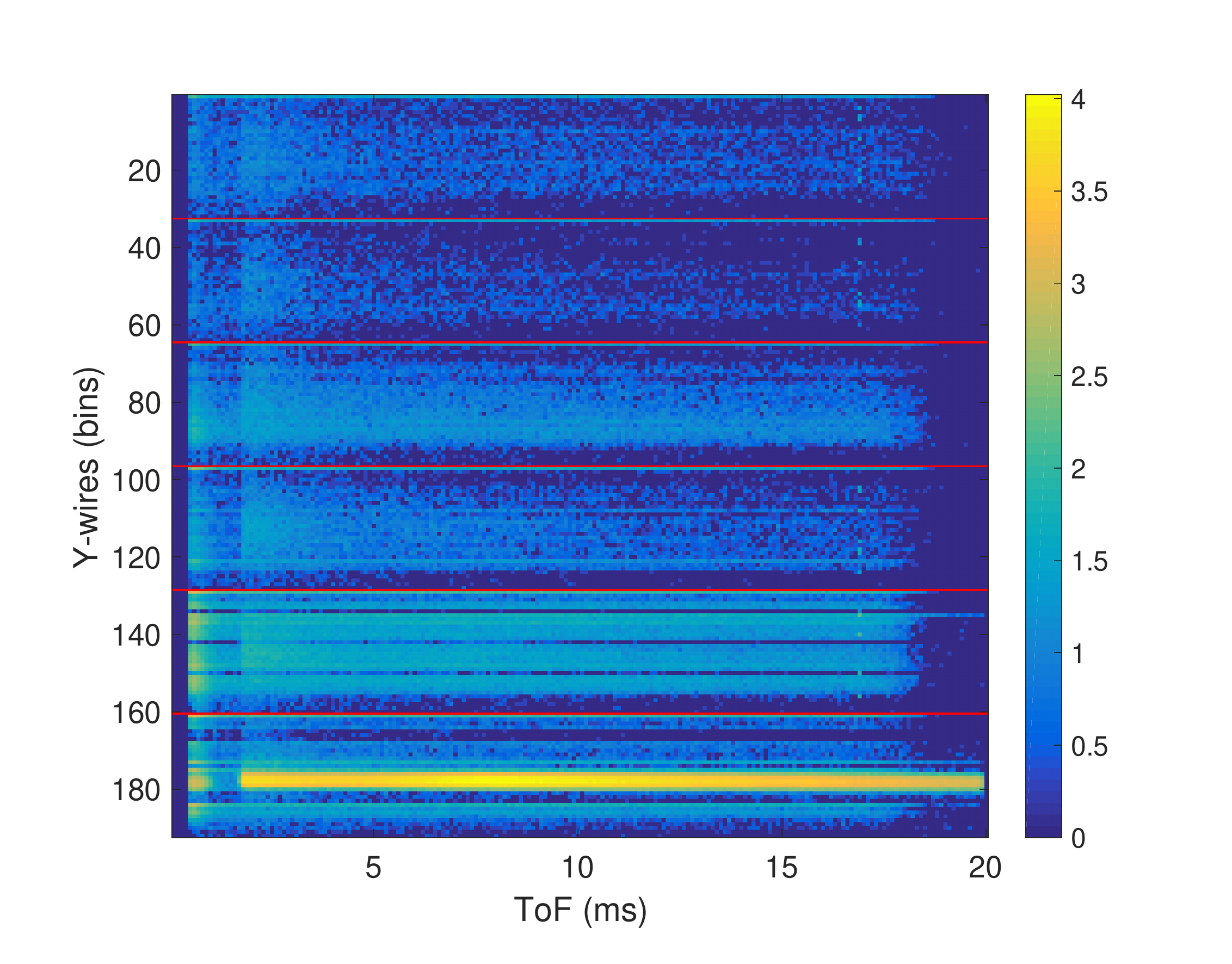}
\caption{\label{fig5} \footnotesize 2D image of six cassettes of the Multi-Blade detector (left). ToF image of the detector integrated over the X-direction (strips) (right). This is a raw uncorrected data with no software threshold applied. The color bar represents counts in logarithmic scale.}
\end{figure}

The hardware thresholds are applied to the individual channels of the digitizers (either wires or strips) and it is needed to reject the electronic noise. Note that these values have been set not to fully discriminate against low energy events, e.g. background $\gamma$-rays. Therefore, a software threshold can be applied to filter these events directly in the data. 
\\ The triplets $(X,Y,ToF)$, that identify an event and were described in section~\ref{MBtes}, can be represented by two-dimensional plots: the 2D image of the detector which corresponds to the $(X,Y)$ coordinates and the ToF image of the detector which corresponds to the $(Y,ToF)$ coordinates and it is integrated over the other spatial coordinate ($X$, the strips). The $(X,ToF)$ image can also be used but it is not relevant for our purposes. Moreover, the 2D image $(X,Y)$ can be either integrated over the ToF coordinate or gated in any range of time. A ToF of $6\,ms$ corresponds approximately to 1.8\AA, $8\,ms$ to 2.5\AA\, and $12.5ms$ to 4\AA.  
\\ Figure~\ref{fig5} shows the 2D image and the ToF image of the part of the detector that has been read out by the six digitizers (six cassettes); the horizontal red lines indicate where each cassette starts and ends. The direct beam was directed to the lower cassette of the Multi-Blade detector without being reflected by any sample and its footprint, at the detector, was approximately $3\,mm \times 60\,mm$. The instantaneous peak rate (as defined in~\cite{DET_rates}) in the beam was of $\approx10^4\,Hz$ (corresponding to an instantaneous local rate of $\approx50\,Hz/mm^2$ at peak).
\\ Note that if no software thresholds are applied nor the wires are requested to be in coincidence with the strips, a constant background is visible and it is mostly due to $\gamma$-rays. This background almost disappear if the coincidences are selected, although no software thresholds are applied. Moreover, the bright peak at $1\,ms$ in ToF has to be attributed to fast and/or epithermal neutrons of wavelength below 0.3\AA . The peak disappears if only events below the software threshold are selected because fast neutrons generally release large amount of energy in the gas~\cite{MIO_fastn}.
\begin{figure}[htbp]
\centering
\includegraphics[width=.49\textwidth,keepaspectratio]{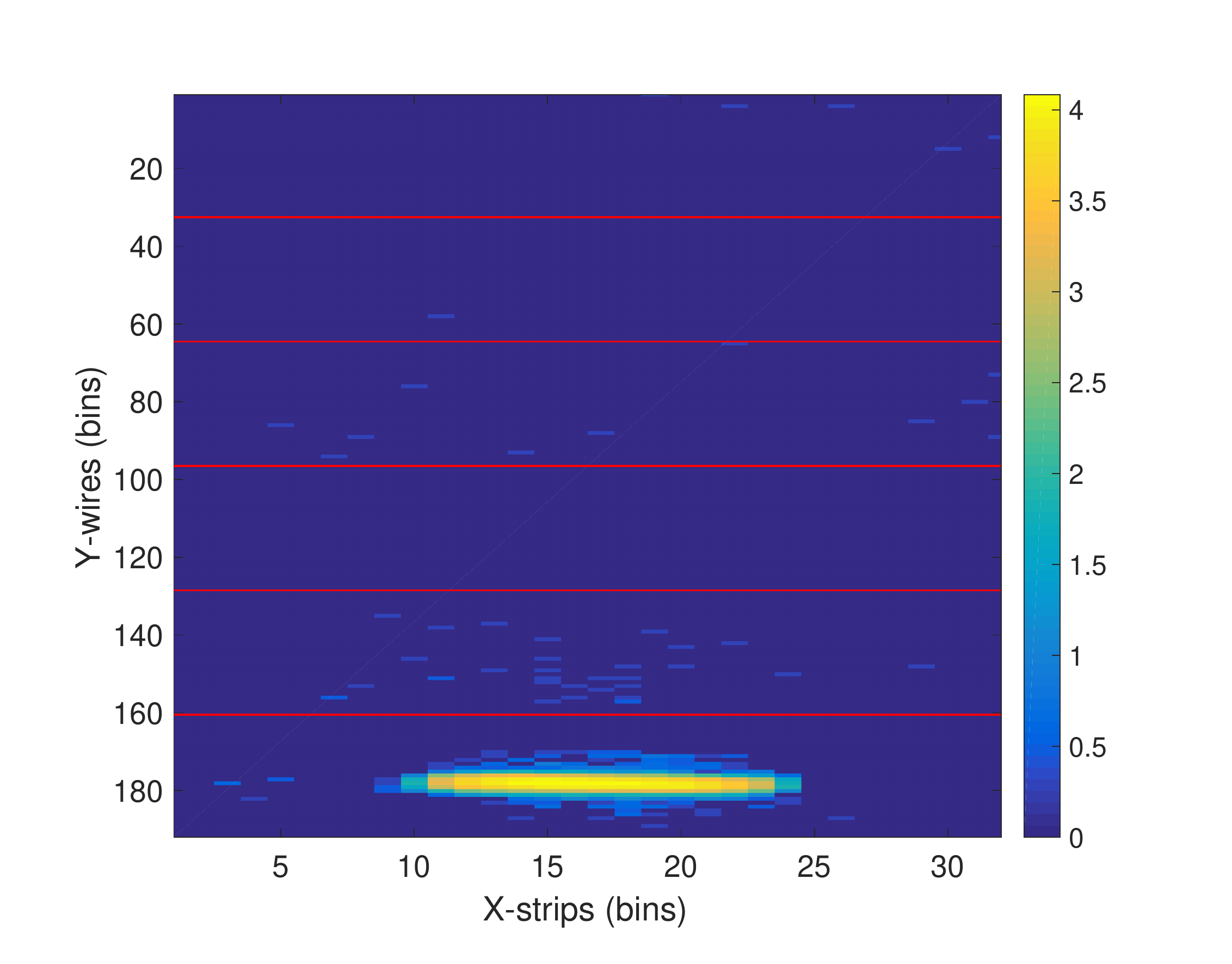}
\includegraphics[width=.49\textwidth,keepaspectratio]{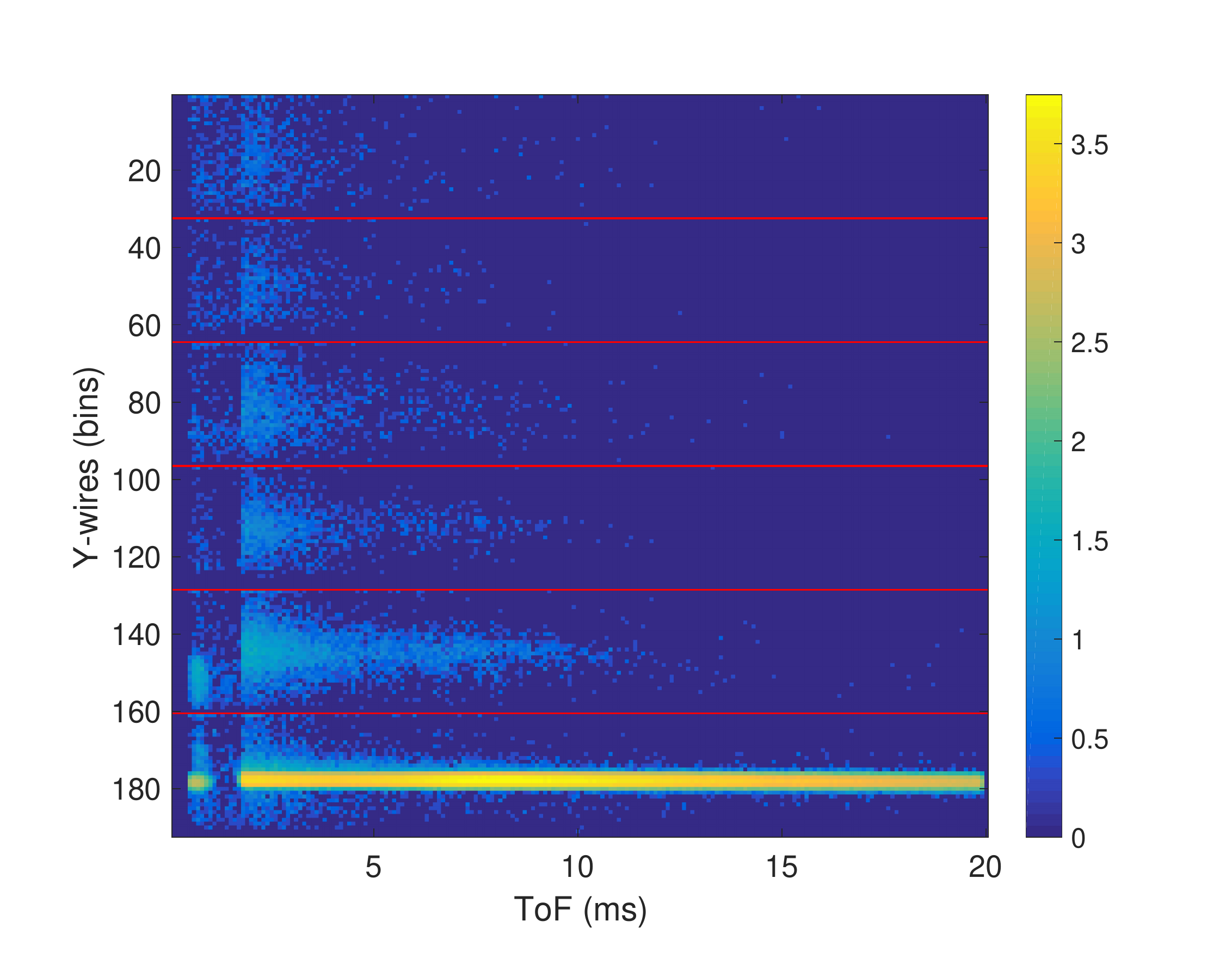}
\caption{\label{fig6} \footnotesize 2D image of the detector when software thresholds are applied to discriminate against $\gamma$-rays and the ToF is gated between $12.5\,ms$ and $20\,ms$ (4\AA\,- 6.5\AA) (left). ToF image of the detector when software threshold is applied (right). The color bar represents counts in logarithmic scale.}
\end{figure} 
\\ When a software threshold is applied, the plot on the right in Figure~\ref{fig6} is obtained. The constant background, which can be attributed to the $\gamma$-rays, vanishes. However a background at short wavelengths ($<4$\AA, $12.5\,ms$) is still visible and reproduces the shape of the direct beam. The plot on the left in Figure~\ref{fig6} depicts the 2D image of the direct beam with software thresholds applied and with the ToF gated between $12.5ms$ and $20\,ms$ (above 4\AA). Note that if the ToF is not selected in the indicated range, the background does not vanish.  
\\ Figure~\ref{fig7} shows the normalized counts in the 6 cassettes from the 2D image in Figure~\ref{fig6} integrated over the X-direction (strips) for four different gates in ToF, below $1.5\,ms$ ($\approx 0.5$\AA), between $1.5\,ms$ and $8\,ms$ (0.5 \AA\,- 2.5\AA), between $8\,ms$ and $12.5\,ms$ (2.5 \AA\,- 4\AA) and between $12.5\,ms$ and $20\,ms$ (4 \AA\,- 6.5\AA). 
\begin{figure}[htbp]
\centering
\includegraphics[width=.7\textwidth,keepaspectratio]{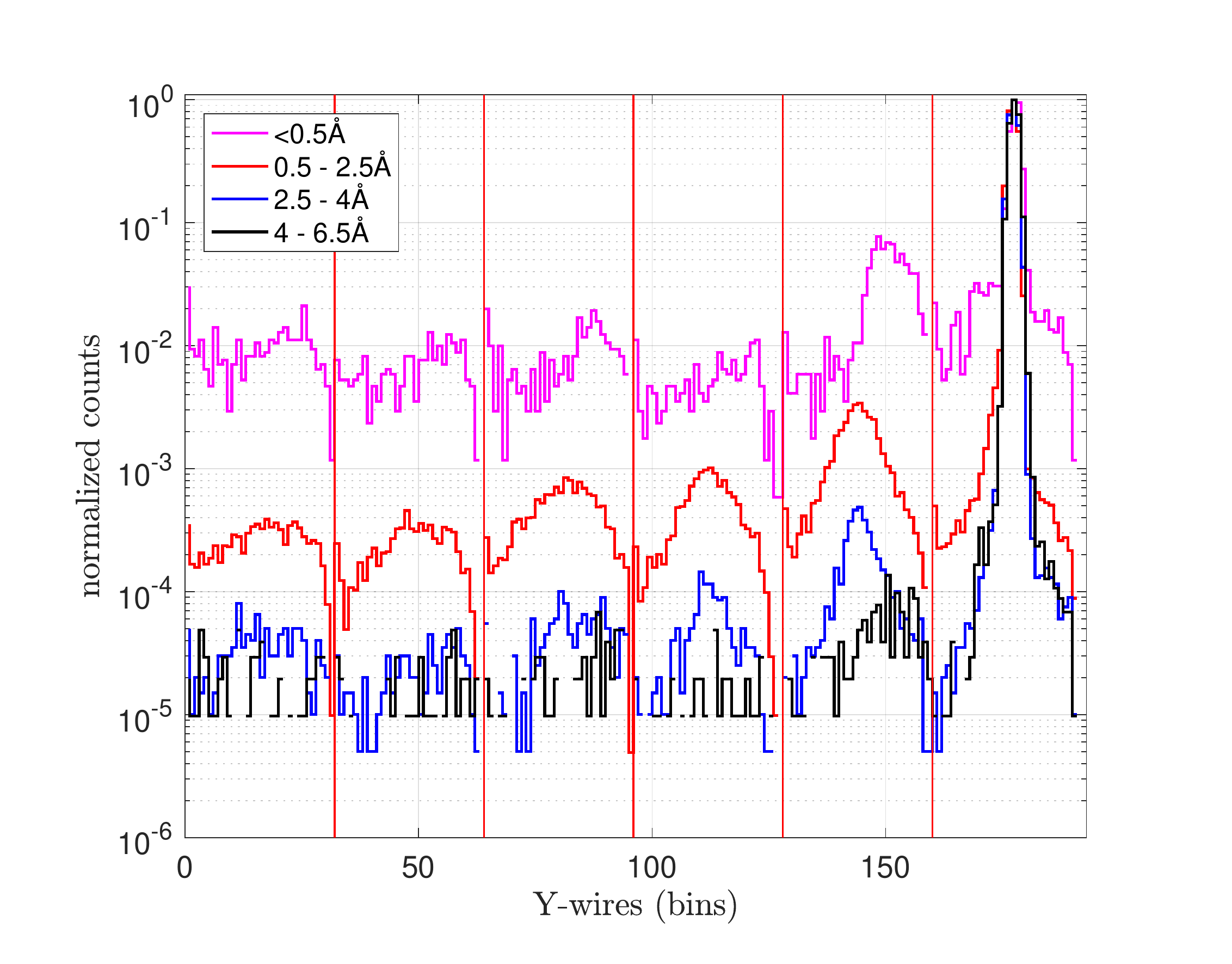}
\caption{\label{fig7} \footnotesize Normalized counts in the 6 cassettes from the 2D image in Figure~\ref{fig6} integrated over the X-direction (strips) for four different gates in the ToF: below $1.5\,ms$ ($\approx 0.5$\AA), between $1.5\,ms$ and $8\,ms$ (0.5 \AA\,- 2.5\AA), between $8\,ms$ and $12.5\,ms$ (2.5 \AA\,- 4\AA) and between $12.5\,ms$ and $20\,ms$ (4 \AA\,- 6.5\AA).}
\end{figure} 
\\ The shape of the direct beam, centered in the lower cassette, is reproduced into the others and the intensity decreases with the distance. This effect can be attributed to the neutrons that cross the $\mathrm{^{10}B_4C}$ coating without being absorbed. They are scattered by the substrate and they are detected in other cassettes. Note that the nominal recommended thickness is $7.5\,\mu m$, but the present blades (Ti and SS) have been coated with $4.4\,\mu m$. Figure~\ref{fig2} shows the amount of absorption in the $\mathrm{^{10}B_4C}$ layer on the blades as a function of the neutron wavelength. 
\begin{figure}[htbp]
\centering
\includegraphics[width=.7\textwidth,keepaspectratio]{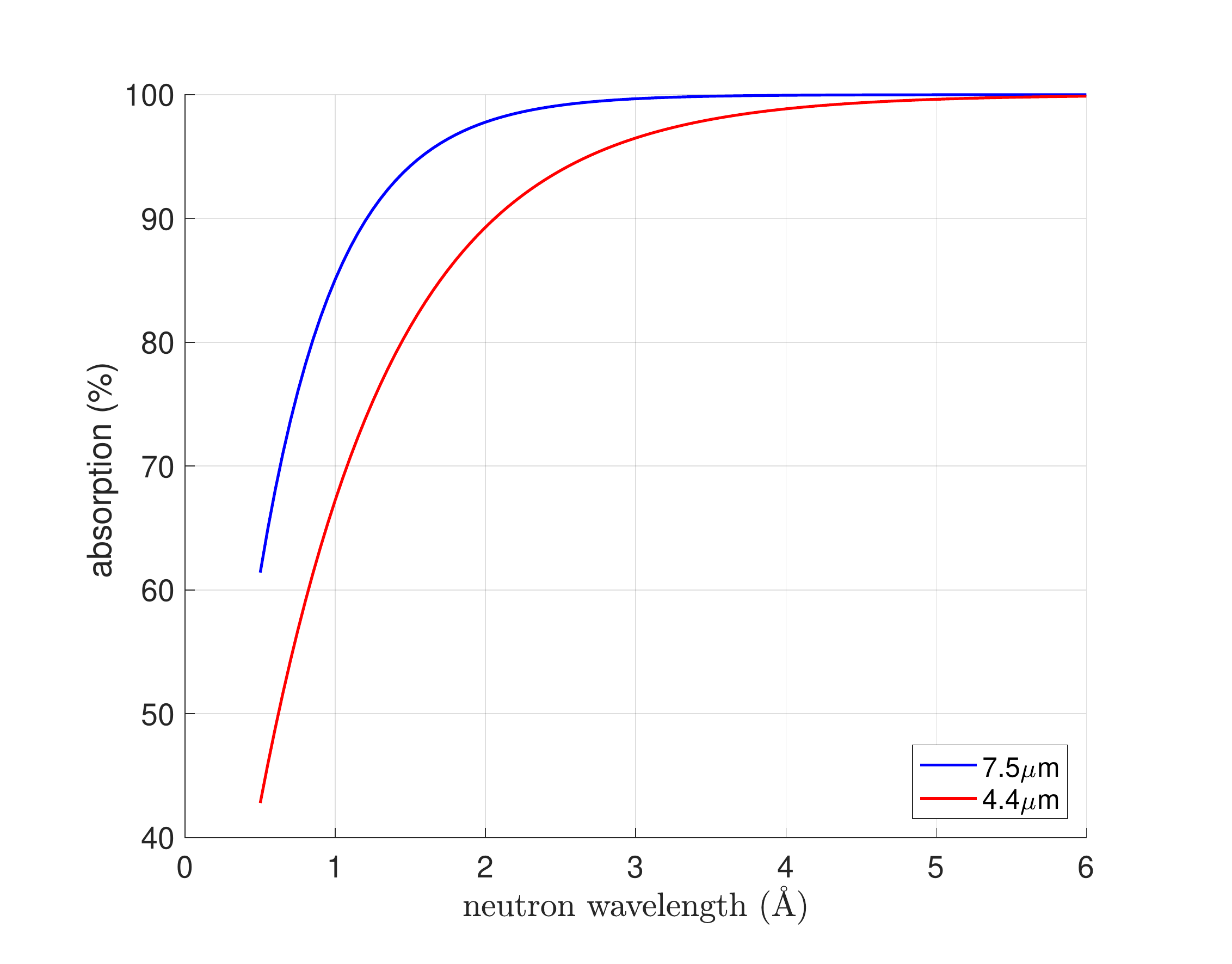}
\caption{\label{fig2} \footnotesize Calculation of the absorbed neutron flux as a function of wavelength for the nominal coating ($7.5\,\mu m$) and for the coating thickness in the present detector ($4.4\,\mu m$) inclined at 5 degrees.}
\end{figure} 
We expect, with the actual coating, that about $50\%$ of neutrons at the shortest wavelengths (see Figure~\ref{fig2}) are not stopped by the layer causing this background. Note that any material chosen among SS, Ti or Al for the substrate of a blade at 5 degrees correspond to about a $11$ times thicker layer. Therefore, the amount of scattered neutron flux from any $2\,mm$-thick substrate of a blade is close to unity~\cite{MIO_MB2017}. 
\\ By applying the gate above $12.5\,ms$ (4\AA) the background is completely suppressed, indeed the absorption efficiency of the $\mathrm{^{10}B_4C}$-coating is always above $98\%$ for this or longer wavelengths. 
\\ From the detector requirements set by the instruments, the shortest wavelength that will be used is 2.5\AA\,(Table~\ref{tab1}). We expect the nominal coating ($7.5\, \mu m$) at 2.5\AA\, to be as efficient at absorbing neutrons as the $4.4\,\mu m$ coating at 4\AA.
\subsection{Dynamic range}
\begin{figure}[htbp]
\centering
\includegraphics[width=.49\textwidth,keepaspectratio]{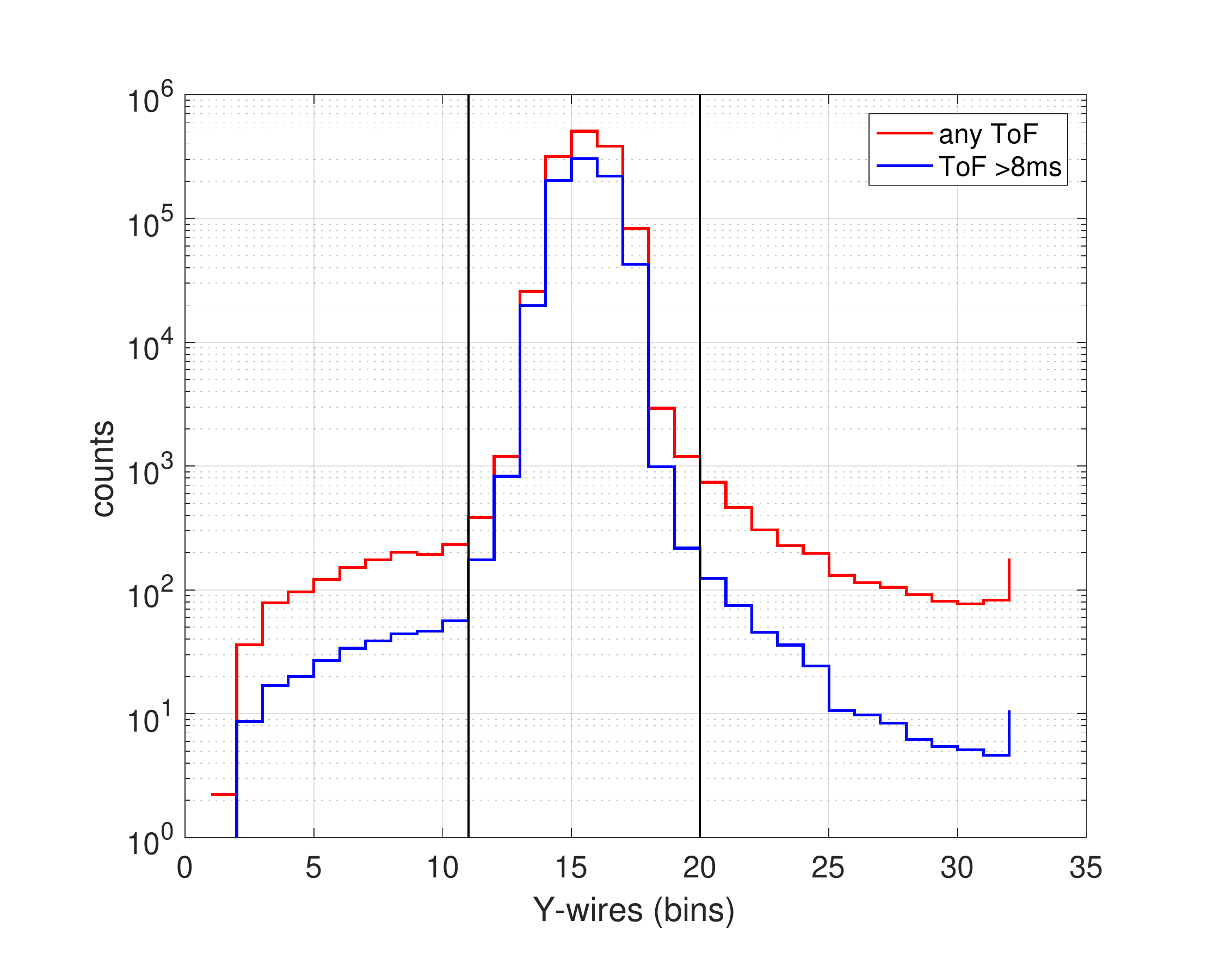}
\includegraphics[width=.49\textwidth,keepaspectratio]{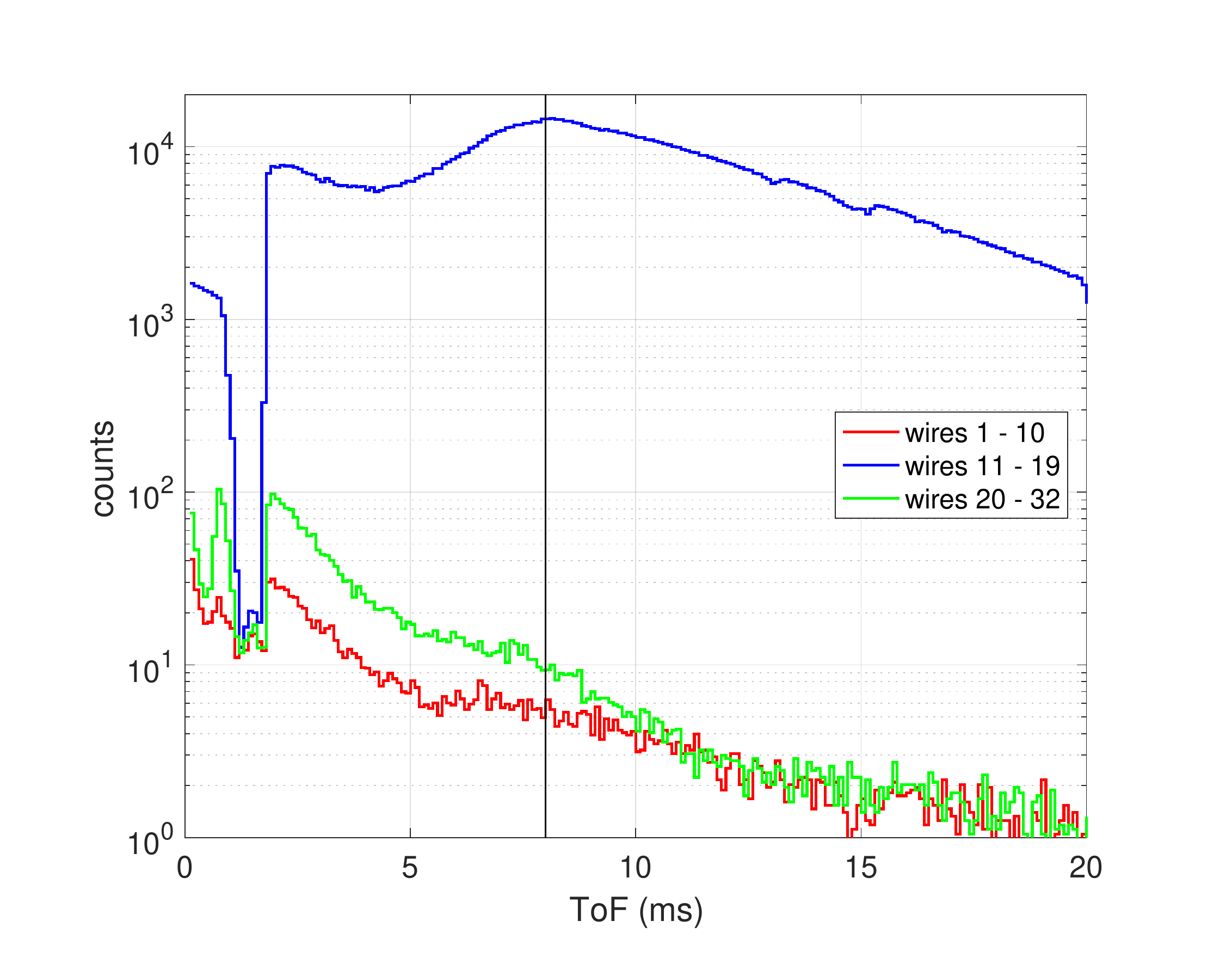}
\caption{\label{fig7bis} \footnotesize Direct beam profile on wires integrated over the strips for the full ToF spectrum and for events gated above $8\, ms$ (2.5\AA) (left). ToF spectra for wires in the direct beam (11-19) and for the wires in the tails of the direct beam (1-10 and 20-32) (right).}
\end{figure}

In order to quantify the spatial and time dynamic range of the detector the direct beam was directed to the lower cassette of the Multi-Blade. The spatial dynamic range is related to the ability of the detector to measure in different pixels (adjacent or not) two different counting rates at the same time: the difference between the counting rates defines the actual spatial dynamic range of the detector. Equivalently, the time dynamic range defines the ability of the detector to measure in the same pixel two different counting rates in subsequent time bins.
\\ Figure~\ref{fig7bis} shows the profile of the direct beam on the wires of the illuminated cassette, integrated over the strips and the relative ToF spectra. The direct beam is comprised within wire no. 11 and no. 19. The tails of the beam extend all over the wire plane. We show the comparison of the beam profiles integrated over the full ToF spectrum and gated above $8\, ms$ (2.5\AA) in order to decrease the contribution of the scattered neutrons within the detector as described in the previous section. The spatial dynamic range is about $10^4$ (peak to tail).
\\ The ToF spectra in Figure~\ref{fig7bis} is shown for the wires in the beam and for those in the tails. Above $8\, ms$, a difference of approximately 3 orders of magnitude in counting rate, in the same ToF bin, is visible between any wire in the interval 11-19 and any other in the tails (1-10 or 20-32). Moreover, the ToF spectrum varies of 3 orders of magnitude within two subsequent time bins (at $\approx 2\, ms$).  
\\ The measured dynamic range with the Multi-Blade detector is the actual dynamic range of the CRISP instrument~\cite{INSTR_OSMOND_CRISP}, a lower background environment would be required to determine the limits of the detector technology.
\subsection{Detection efficiency}
\begin{figure}[htbp]
\centering
\includegraphics[width=.49\textwidth,keepaspectratio]{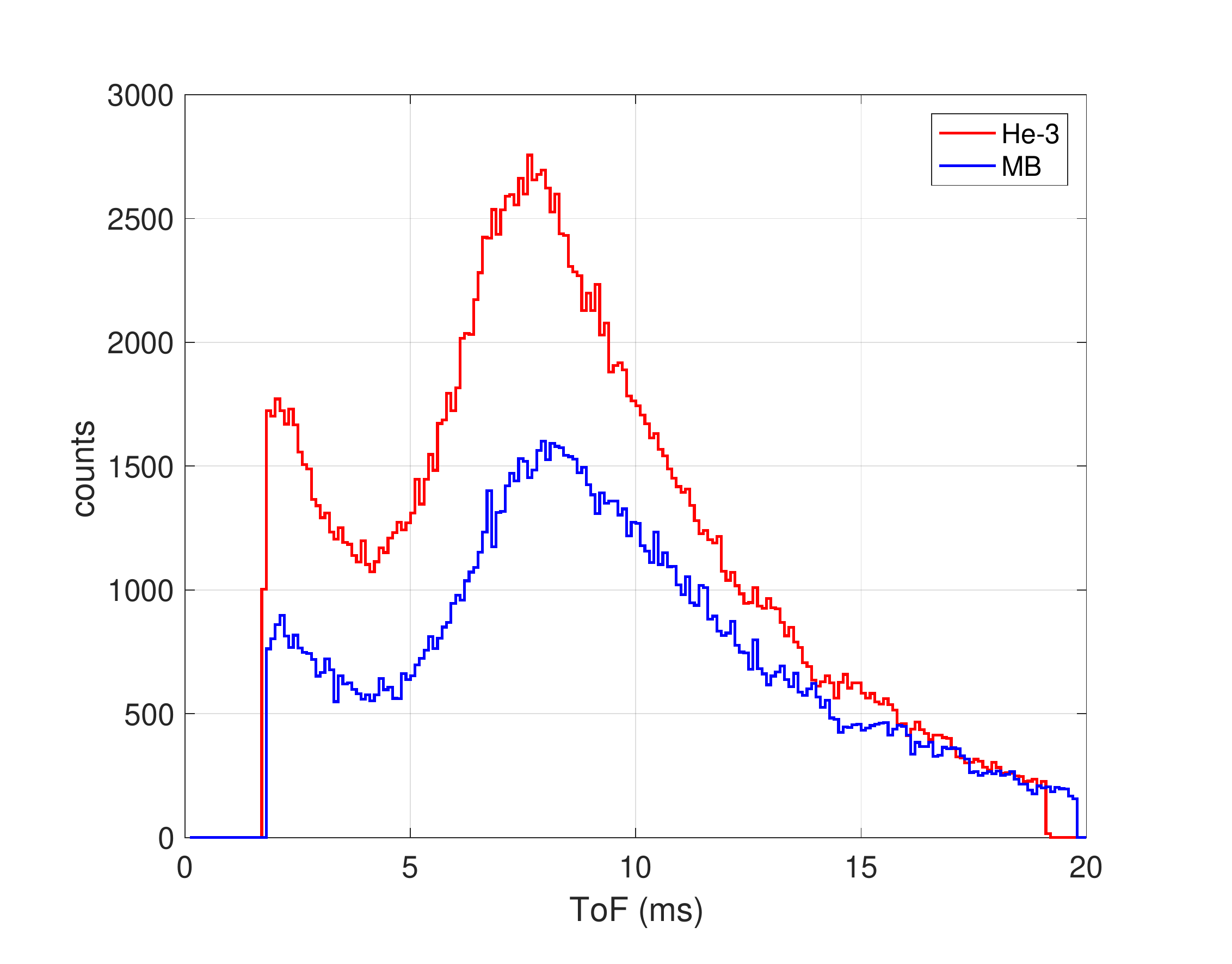}
\includegraphics[width=.49\textwidth,keepaspectratio]{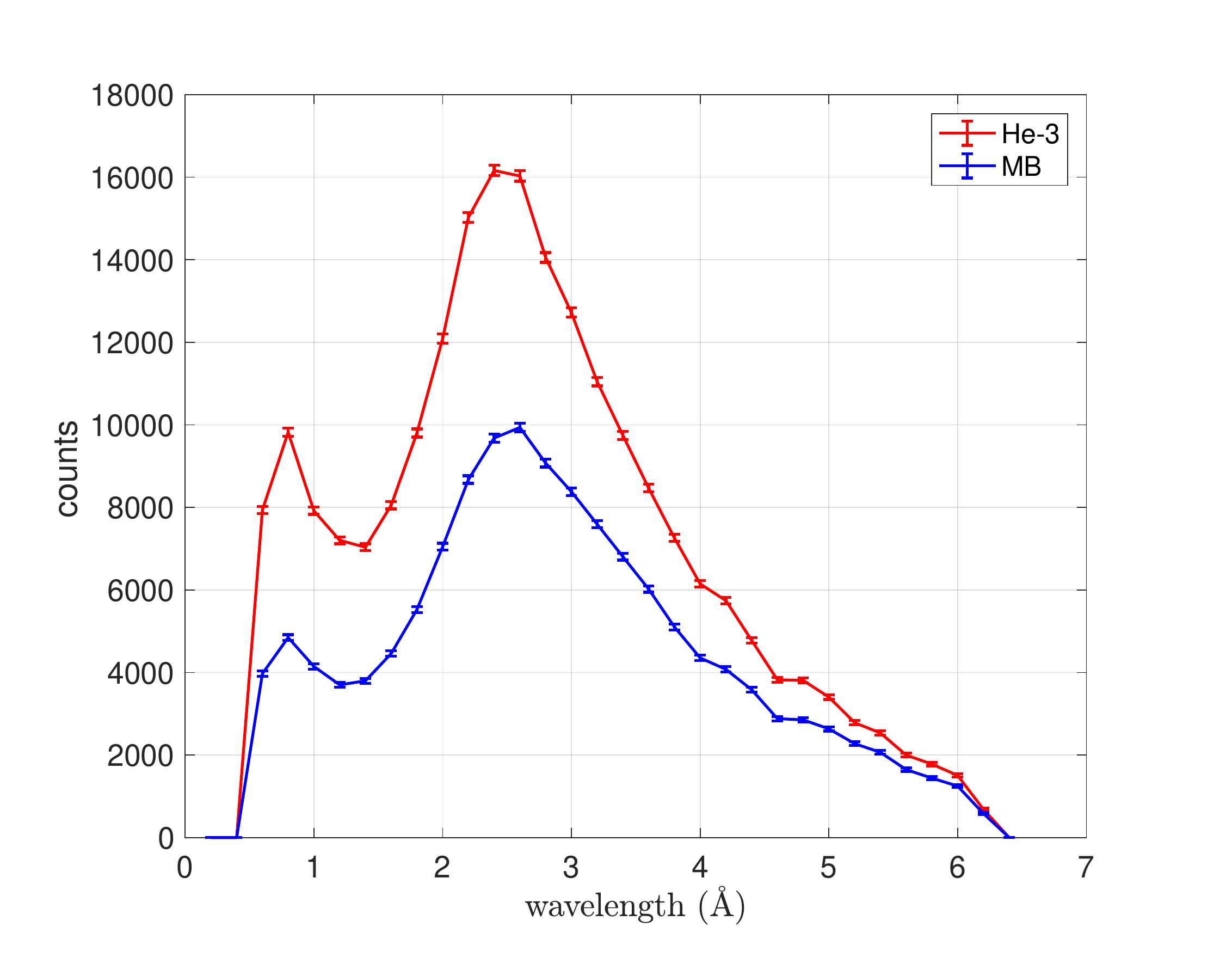}
\caption{\label{fig8} \footnotesize The raw uncorrected ToF spectra of the CRISP $\mathrm{^3He}$ detector and the Multi-Blade (left). The difference in the ToF spectrum of $0.5\,ms$ for the slower neutrons is due to the fact that the $\mathrm{^3He}$ detector was $\approx 0.5\,m$ closer to the sample than the Multi-Blade detector. The spectra of the CRISP $\mathrm{^3He}$ detector and the Multi-Blade as a function of the neutron wavelength (right). Note that the ToF spectrum of the Multi-Blade detector is corrected with the depth of the detector according to the equation~\ref{equadep}.}
\end{figure}

\begin{figure}[htbp]
\centering
\includegraphics[width=.49\textwidth,keepaspectratio]{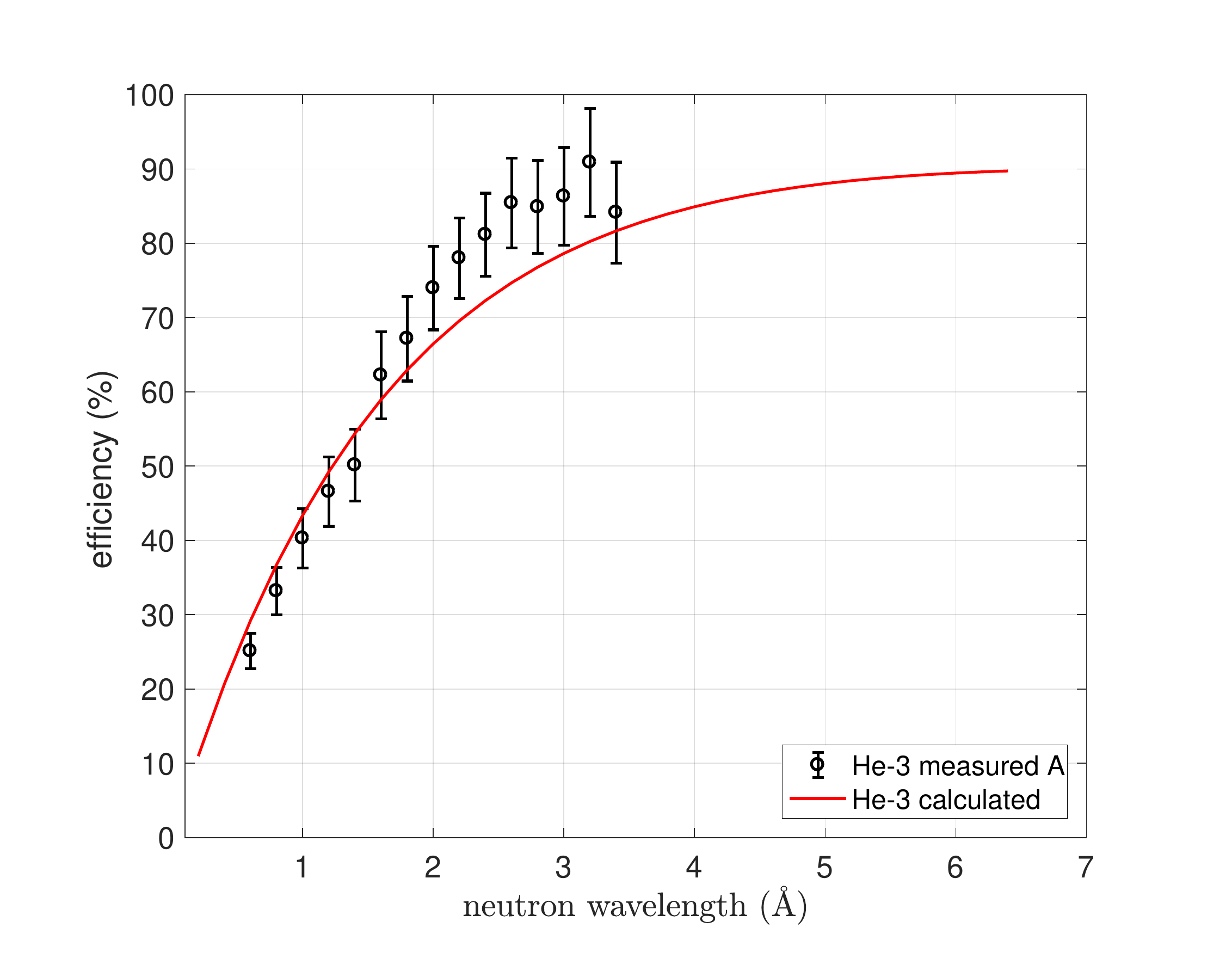}
\includegraphics[width=.49\textwidth,keepaspectratio]{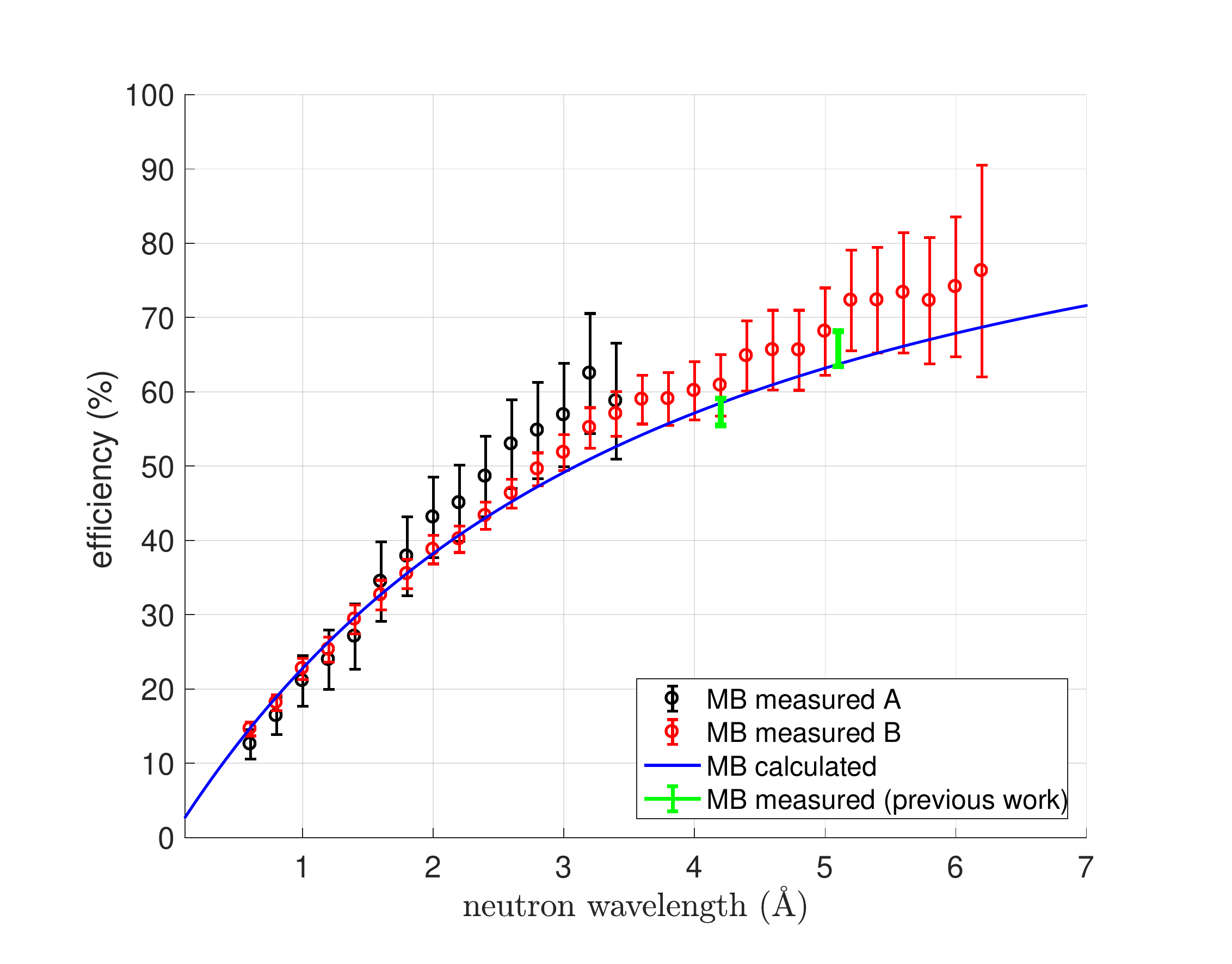}
\caption{\label{fig8bis} \footnotesize The measured and the calculated efficiency of the $\mathrm{^3He}$ detector on CRISP (left). The measured efficiency can be considered valid up to 3.5\AA\, ($11\,ms$). The efficiency is calculated assuming a $\mathrm{^3He}$ gas pressure of $3.5\,bar$. The efficiency of the Multi-Blade detector calculated according to~\cite{MIO_analyt} and measured against the $\mathrm{^3He}$ tube of CRISP (right). The efficiency is normalized assuming either the measured or calculated efficiency of the $\mathrm{^3He}$ detector, measurement A and measurement B respectively. The two measured point from a previous detector charaterization~\cite{MIO_MB2017} are also shown.}
\end{figure}

The detection efficiency is defined as the ratio of detected neutrons over the incoming neutrons in the beam in a defined area. A set of data is recorded with the Multi-Blade detector using the direct beam. Software thresholds were applied to the data in order to discriminate against background events. The same configuration was used to illuminate the $\mathrm{^3He}$-detector of CRISP which was previously calibrated, thus its efficiency, as a function of the neutron wavelength, is known. The data for the $\mathrm{^3He}$-tube efficiency can be considered valid up to 3.5\AA\, ($11\,ms$) due to high background at larger wavelengths. Figure~\ref{fig8bis} (left) shows the measured efficiency of the CRISP detector up to 3.5\AA\,and the calculated efficiency for a $\mathrm{^3He}$ gas pressure of 3.5 bar. Since the absolute efficiency of the $\mathrm{^3He}$-detector is known, the absolute efficiency of the Multi-Blade can be calculated. The ratio between the ToF spectra of the two detectors defines their relative efficiency. This method is complementary and independent to the more commonly used method employing monochromatic pencil beams and it has different and independent systematic effects affecting the uncertainty on the final result.
\\ We integrate the counts in the beam over the spatial coordinates of the Multi-Blade and we show the resulting ToF spectrum in Figure~\ref{fig8} (left) along with the spectrum of the $\mathrm{^3He}$-tube. The $\mathrm{^3He}$ is physically $0.5\,m$ closer to the sample than the Multi-Blade, this results into a difference in the ToF spectrum, i.e. the slower neutrons above 6\AA \, ($\approx 20\,ms$), arrive at the Multi-Blade detector with $0.5\,ms$ delay with respect to the $\mathrm{^3He}$ tube. Hence, the two ToF spectra are slightly shifted and stretched relative to each other. Both the ToF spectra can be plotted as a function of the neutron wavelength by knowing the distance from the target: $12.6\,m$ and $12.1\,m$ for the Multi-Blade and the $\mathrm{^3He}$-tube respectively. Figure~\ref{fig8} (right) shows the spectra of the two detectors as a function of the neutron wavelength. Observe that the ToF spectrum of the Multi-Blade detector is corrected with the depth of the detector according to the equation~\ref{equadep}. 
\\ The ratio between the Multi-Blade and $\mathrm{^3He}$ detector spectra as a function of the neutron wavelength can be calculated and this gives us the relative efficiency of the two detectors. In order to calculate the absolute Multi-Blade efficiency the latter must be normalized to the absolute efficiency of the  $\mathrm{^3He}$ detector. The absolute Multi-Blade detector efficiency is shown in Figure~\ref{fig8bis} (right) as a function of the neutron wavelength and compared to the theoretical efficiency calculated according to~\cite{MIO_analyt,MIO_decal}. The black points (measurement A) represent the calculated efficiency for the Multi-Blade by normalizing to the measured efficiency of the $\mathrm{^3He}$ detector, whereas the red points (measurement B) are the calculated efficiency for the Multi-Blade by normalizing to the $\mathrm{^3He}$ detector efficiency calculated from the $\mathrm{^3He}$ pressure ($3.5\,bar$). 
\\ The measured efficiency shifts systematically toward higher values at larger wavelengths, this is also visible in the measured $\mathrm{^3He}$ detector efficiency which propagates in the Multi-Blade efficiency normalization. The overall trend is as expected; the measurements are consistent with calculation~\cite{MIO_analyt}, with the exception of the points at the longest wavelength, where the Multi-Blade are slightly above expectation. Moreover, the obtained Multi-Blade efficiency agrees with the previously measured efficiency shown in~\cite{MIO_MB2017}. 
\subsection{Stability}
\begin{figure}[htbp]
\centering
\includegraphics[width=1\textwidth,keepaspectratio]{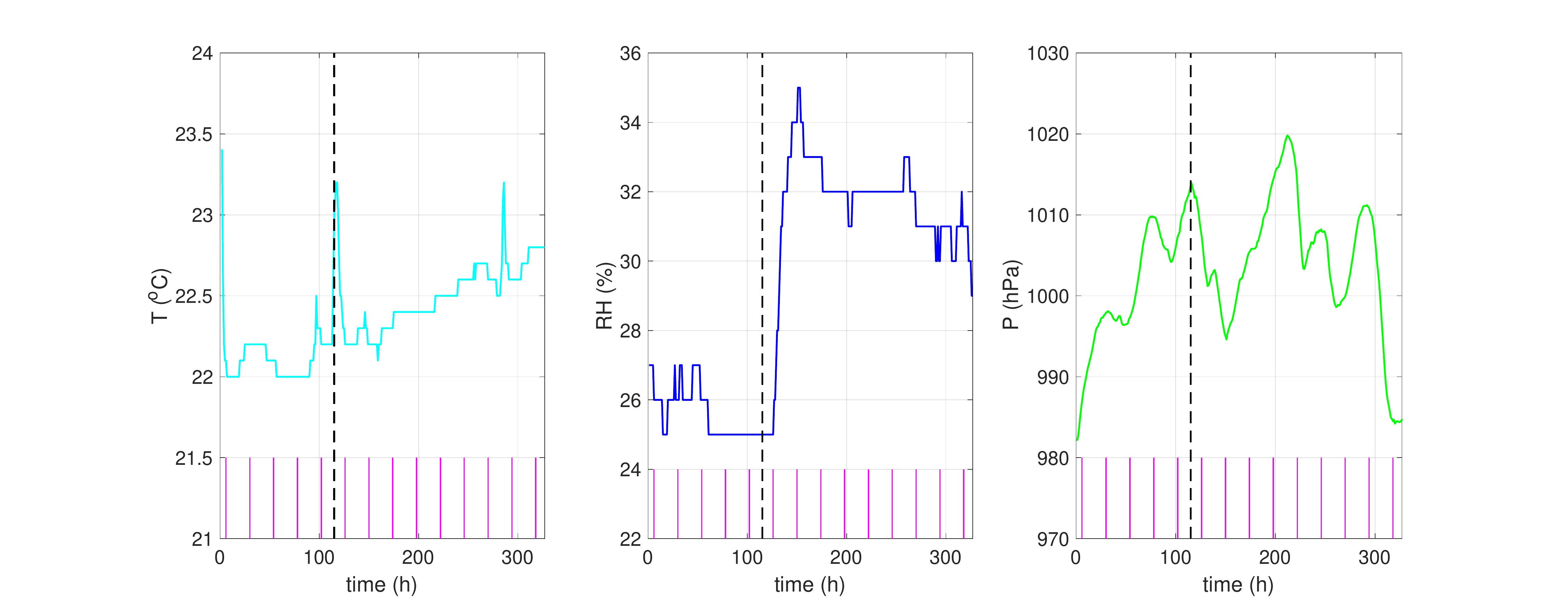}
\caption{\label{fig9bis} \footnotesize Trends of the recorded temperature (left), relative humidity (centre) and atmospheric pressure (right). The pink vertical lines indicate the midnight of each day. The vertical black dashed line indicates the change of the detector flow from $60\,l/day$ to $30\,l/day$.}
\end{figure}

\begin{figure}[htbp]
\centering
\includegraphics[width=1\textwidth,keepaspectratio]{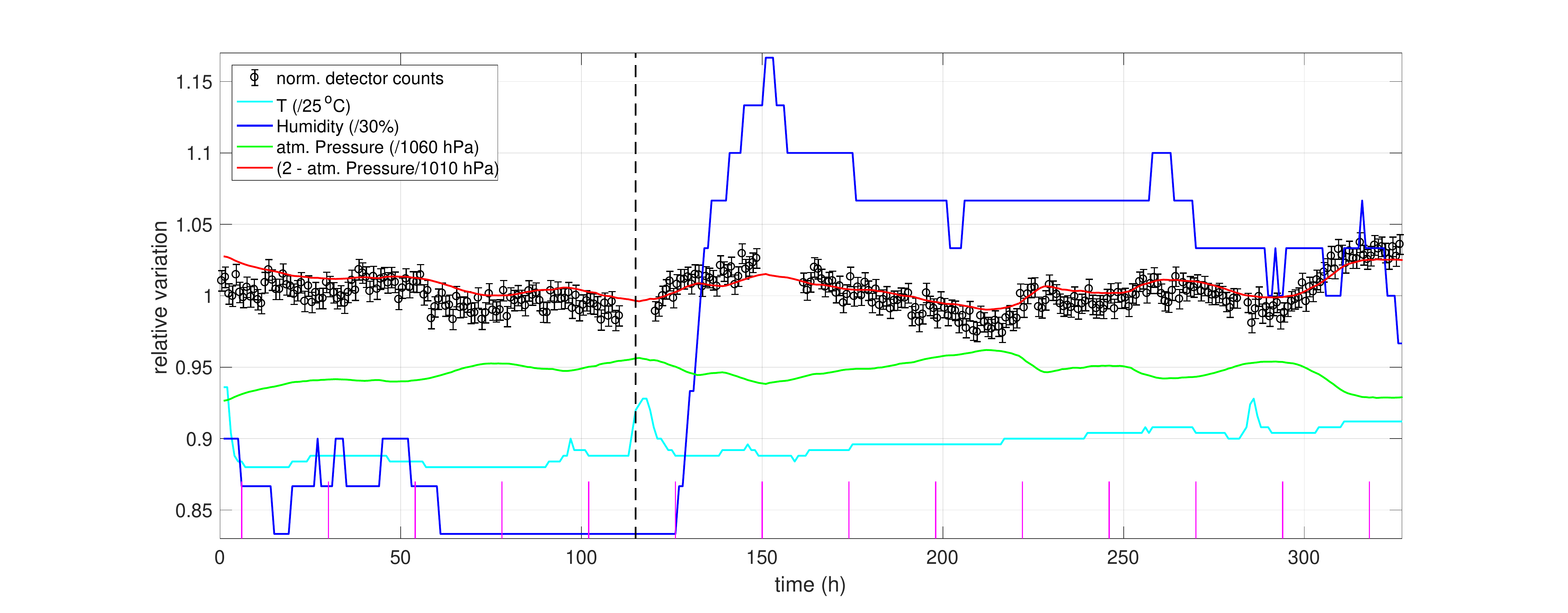}
\caption{\label{fig9} \footnotesize Normalized detector counts over 2 weeks along with the temperature, humidity and atmospheric pressure (also normalized). The atmospheric pressure is also shown as $(2-P/1010hPa)$. The pink vertical lines indicate the midnight of each day. The vertical black dashed line indicates the change of the detector flow from $60\,l/day$ to $30\,l/day$. }
\end{figure}

The gas gain of a MWPC is well described by the Diethorn's formula~\cite{DET_Stability_Diet} and it is influenced by the atmospheric pressure and temperature variation as described in~\cite{DET_Stability_Bloom,DET_Stability}. 
\\ The Multi-Blade detector was placed in front of a moderated $\mathrm{Pu/Be}$ neutron source at the Source Testing Facility (STF)~\cite{SF2,SF1} at the Lund University in Sweden. The detector was flushed with $\mathrm{Ar/CO_2}$ (80/20) at $2.4\,l/h$ resulting into approximately $60\,l/day$. Since the detector vessel is approximately $30\,l$, the full gas volume is renewed twice per day. A set of data is recorded every hour for approximately two weeks. The total number of counts in the detector integrated over an hour, normalized to the average counting rate, is shown in Figure~\ref{fig9} as a function of time. The temperature, humidity and atmospheric pressure were also monitored and they are shown in Figure~\ref{fig9bis}. The temperature varied within $\approx0.5\,\%$ ($^oC$), the relative humidity within $\approx30\,\%$ and the atmospheric pressure within $\approx4\,\%$ in the two weeks of the measurements. 
\begin{figure}[htbp]
\centering
\includegraphics[width=.7\textwidth,keepaspectratio]{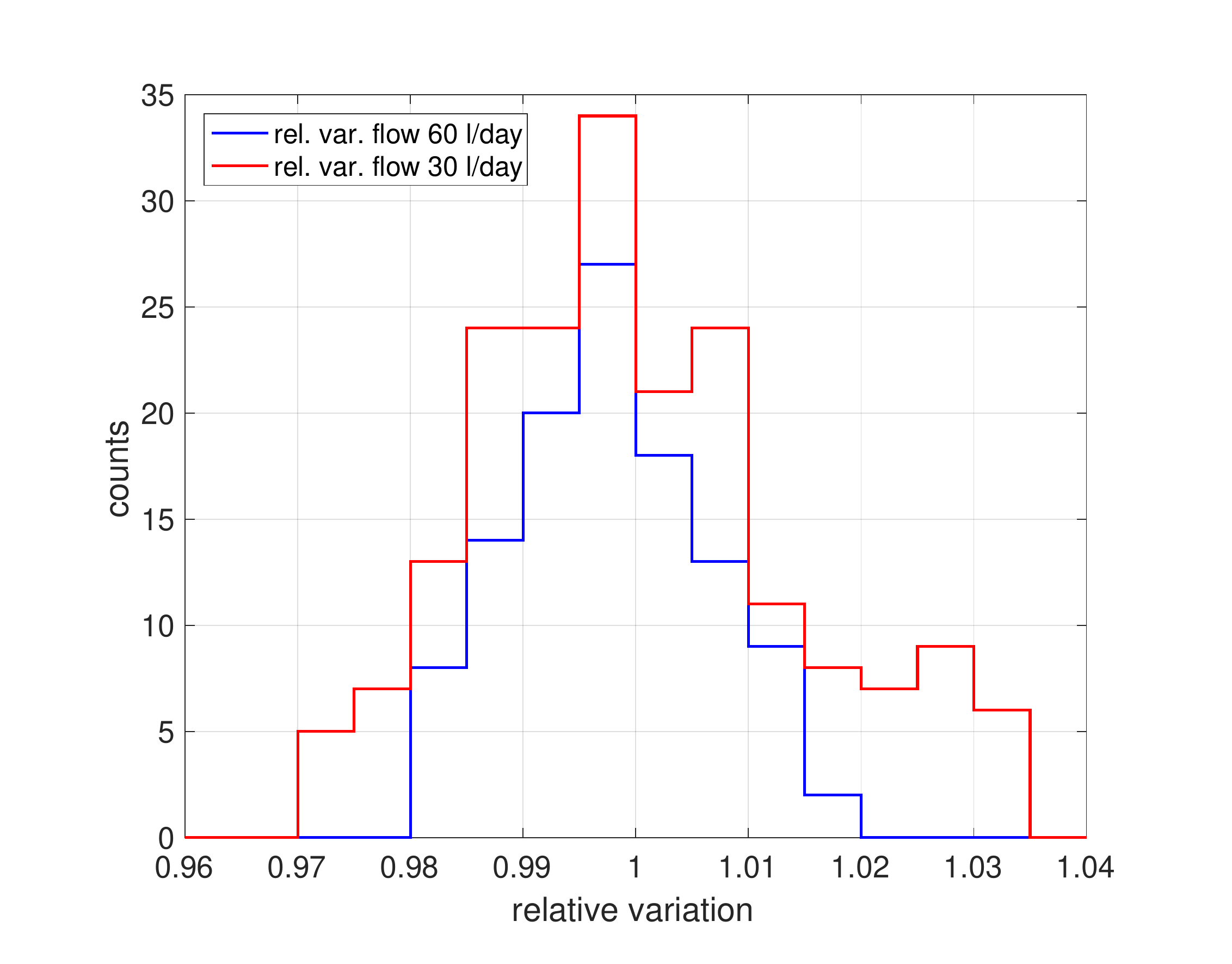}
\caption{\label{fig9bis2} \footnotesize Histogram of the relative variation of the detector counts for the first 115 hours with a gas flow of $60\,l/day$, and for the rest of the measurements with a gas flow of $30\,l/day$.}
\end{figure} 
In Figure~\ref{fig9}, the trend of the atmospheric pressure is also shown as $(2-P/1010hPa)$ in order to better visualize the change of the detector counts with pressure. The missing points in the detector counts corresponds to periods when the Source Testing Facility (STF) was occupied with other users and background was altered, thus we disregard those points. 
\\ After approximately 115 hours of measurement (see the vertical black dashed line in the plot in Figure~\ref{fig9}) the detector flow was changed to one volume per day, i.e. $\approx 1.7\,l/h$ ($30\,l/day$). In the first 115 hours the atmospheric pressure relative variation is $\approx \pm 1.5\%$ and the detector counts variation is $\approx \pm 1.7\%$ (expressed as a range (max-min)/2). Figure~\ref{fig9bis2} shows the histogram of the relative variation of the detector counts for the two gas flow regimes ($60\,l/day$ and $30\,l/day$).
\\ When the flow is reduced, the atmospheric pressure relative variation is $\approx \pm 1.7\%$ and the detector counts variation is $\approx \pm 3\%$. 
\\ In either cases of gas renewal of 1 or 2 volumes per day, we do not observe any contamination of the gas due to pollutant in the detector. A steady decrease of the counts would be visible otherwise.
\\ The trend in detector counts is clearly influenced by the atmospheric pressure. Although the atmospheric pressure variation was comparable in the two configurations of flow, the detector counts are influenced more when the flow is lower. This has to do with the higher over-pressure set in the detector to allow a larger flow. A larger over-pressure in the detector is affected by the variation of the atmospheric pressure to a minor extent. 
\\ The counting rate in the detector is stable within $\pm 1.7\%$ during several days with a flow that replace approximately 2 detector volumes per day. 
\\ In order to further improve the detector stability in time, the gas gain or thresholds must be adjusted according to the atmospheric pressure and temperature variations. Thus, an active feedback on the signal thresholds or on the high voltage as shown in~\cite{DET_Stability} or a post-processing of the data can be used. 
\subsection{Overlap, uniformity and linearity}\label{linea}
\begin{figure}[htbp]
\centering
\includegraphics[width=.8\textwidth,keepaspectratio]{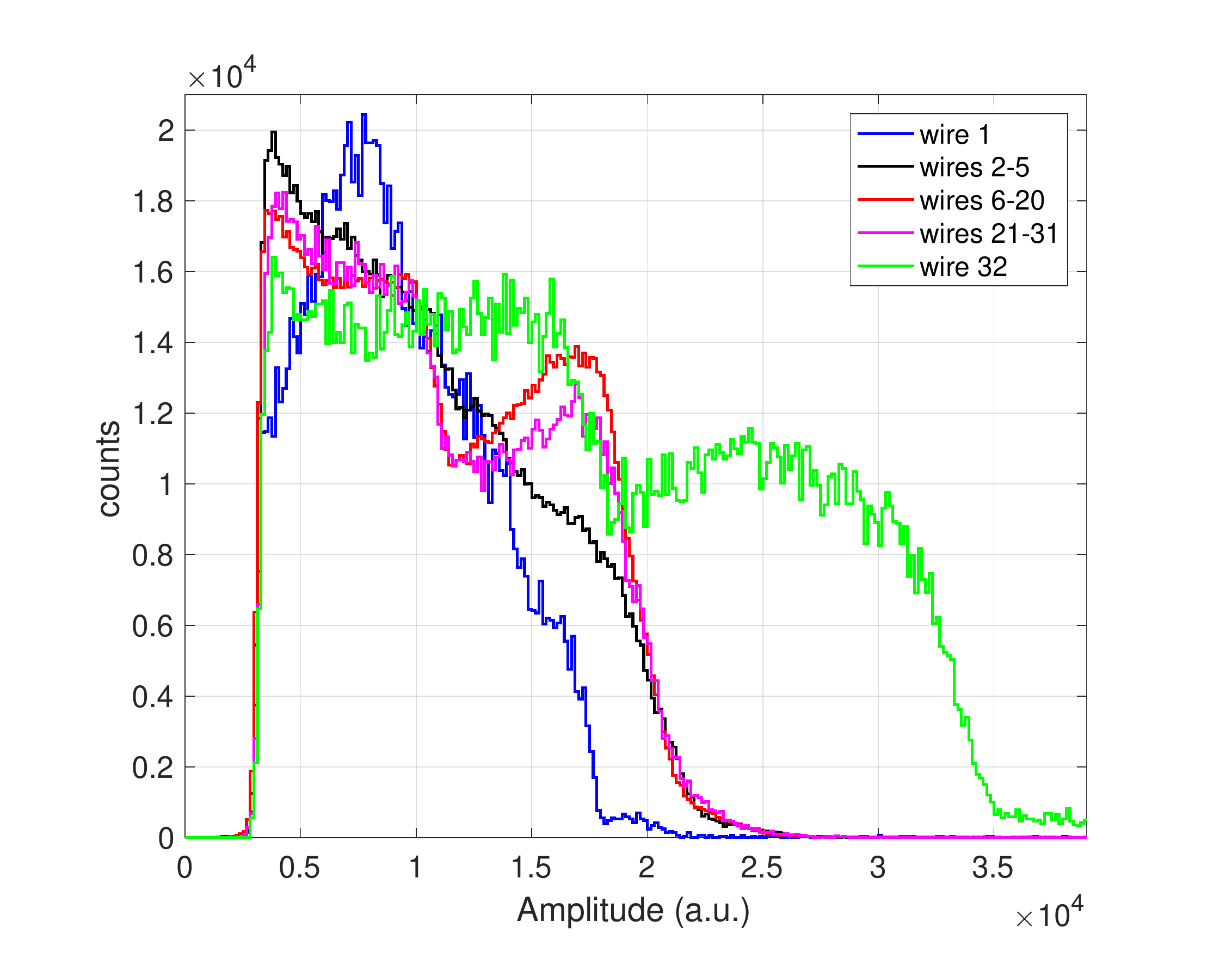}
\caption{\label{fig10} \footnotesize Pulse Height Spectrum (PHS) for individual wires or group of wires in one cassette of the Multi-Blade detector. Wire 1 is at the front of the detector and wire 32 is at the back. The gas gain varies due to the geometry of the cassette.}
\end{figure}

\begin{figure}[htbp]
\centering
\includegraphics[width=.49\textwidth,keepaspectratio]{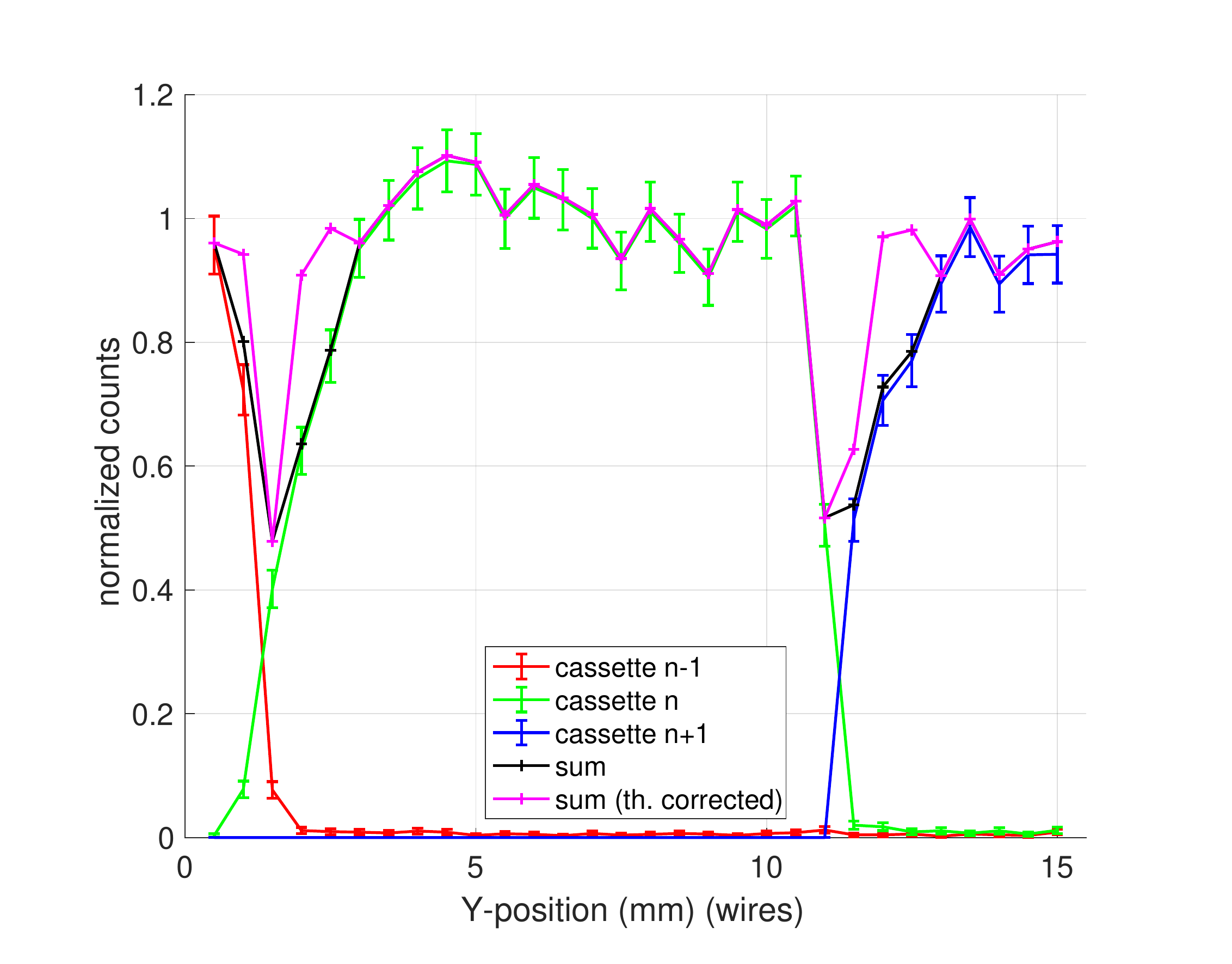}
\includegraphics[width=.49\textwidth,keepaspectratio]{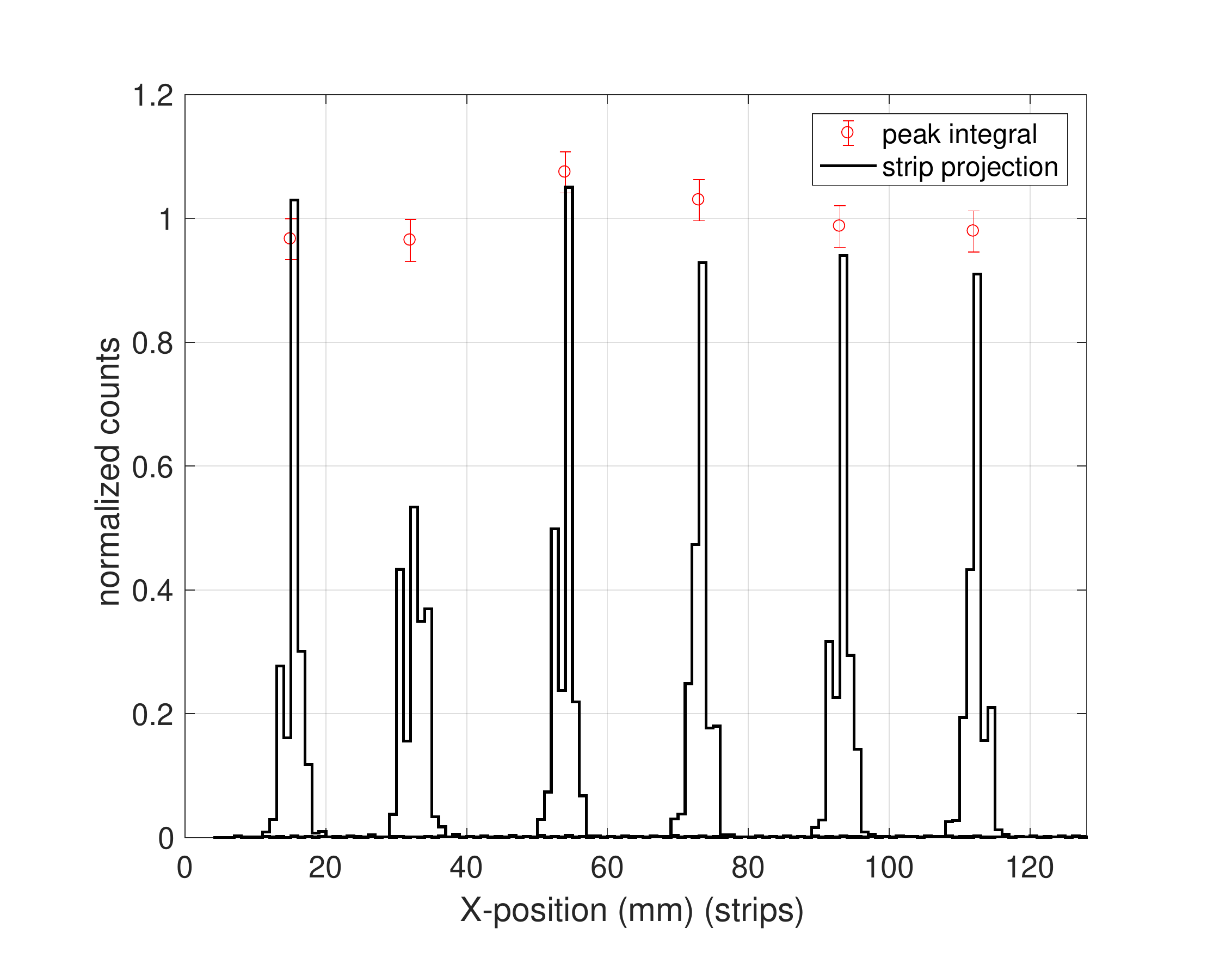}
\caption{\label{fig11} \footnotesize The normalized counts of the Multi-Blade detector scanned across the wires of three adjacent cassettes with a $0.5\,mm$ step (left). In black the sum of the three cassettes and in pink the counts in the three cassettes when the threshold is adjusted for individual channels. The normalized counts of the Multi-Blade detector scanned across the strips of one cassette in steps of $\,20mm$ (right). The red points are the integral of the peaks.}
\end{figure}

Due to the blade geometry the gas gain differs for different wires within a cassette. Each cassette has 32 wires and 32 strips. We label the wire no.1 the one closer to the sample position (see Figure~\ref{fig99}), i.e. at the front of each cassette, and the wire no. 32 the one at the back. Electric field simulations and measurements have been carried out to investigate the gain variation at each wire due to the geometry and they have been described previously in~\cite{MIO_MB2017}. The gain is approximately constant for each wire from no. 8 to no. 31. The wire no. 32 has appropriately a double gain due to the lack of a neighbour. Each of the wires from no. 1 to no. 7 have a different and smaller gas gain with respect to those between no. 8 and no. 31. As shown in~\cite{MIO_MB2017}, the gain drop at the first 7 wires can be compensated by adjusting individual thresholds, in hardware or software, on each channel.
Otherwise the gas gain compensation can be implemented with a separated high voltage supplies at the first wires. Figure~\ref{fig10} shows the Pulse Height Spectrum (PHS) for the wires in a cassette. The peaks from the $\mathrm{n(^{10}B, \alpha)^7Li}$ reaction can be identified for any wire apart from the front wires (from no. 1 to no. 5). It has been shown in~\cite{MIO_MB2017} that even if the threshold is adjusted according to the gain, at the very first wire there is a loss in gain which corresponds to a drop of $50\%$ with respect to the nominal efficiency. This region of reduced sensitivity, is where two cassettes overlap and it is about $0.5\,mm$ wide. 
\\ The Multi-Blade detector was scanned across three cassettes and with a collimated beam. A set of data was recorded for each position in steps of $0.5\,mm$. Figure~\ref{fig11} (left plot) shows the normalized counts of the detector as for each position and for each cassette (red, green and blue curves). The black curve in the plot is the sum over the 3 cassettes with no threshold correction applied: the counts drop due to the gain variation in the two overlap regions scanned. The pink curve is then obtained if the thresholds are adjusted for individual channels and this results in the reduced sensitivity region of about $0.5\,mm$. This meets expectation but can be improved. 
The scan was repeated across the strips in steps of about $20\,mm$. Figure~\ref{fig11} (right plot) shows the normalized counts for each position of the scan and the integral of the counts (also normalized) in each peak. The overall variation of the gain, i.e. the uniformity, in the scanned cassette in both directions (wires and strips) is $\pm 10\%$.
\newpage
\subsection{Masks and reconstruction algorithms}\label{spatresd}
A set of images of BN (Boron-Nitride, HeBoSint C100~\cite{hebosint}) masks were captured with the Multi-Blade on CRISP to investigate how the the position reconstruction algorithm affects the reconstructed image. The BN masks are $5\,mm$ thick and natural enriched in boron. The typical attenuation of these masks is approximately $100\,\%$ from and above 1\AA. Two algorithms can be used to reconstruct the $(X,Y)$ coordinates from the raw data: a maximum amplitude (or area) algorithm (MAX) or a Center of Gravity algorithm (CoG). The first associates the hit of a cluster to the two channels (wire and strip) which have the maximum area (i.e. energy deposition) among the events in the cluster, thus the $X $and $Y$ coordinates are 32 integers for wires and 32 integers for strips. The MAX algorithm does not exploit the information about the multiplicity of an event. On the hand, the CoG algorithm uses the multiplicity in a cluster to better position the hit across the two coordinates. If two, or more, adjacent channels are firing at the same time and belong to the same cluster, the position of the hit is calculated weighting the energies (areas or amplitudes) of the channels. I.e. if two adjacent strips are firing at the same time and they perfectly share the energy in two identical parts, the hit will be placed exactly in the middle of the two. We use 128 bins instead of 32 in the case of CoG. The CoG is an continous quantity and the resolution improves independently from the binning but we choose to bin it in 4 times more bins. In the reconstructed position of the wires there is not much difference if the MAX or the CoG algorithm is used since the multiplicity of wires is often 1. It has been shown that the spatial resolution improves from $\approx0.6\,mm$ to $\approx0.55\,mm$~\cite{MIO_MB2017}. On the other hand, the most common multiplicity on strips is 2, and the CoG algorithm improves significantly the spatial resolution for the strips. 

\begin{figure}[htbp]
\centering
\includegraphics[width=.95\textwidth,keepaspectratio]{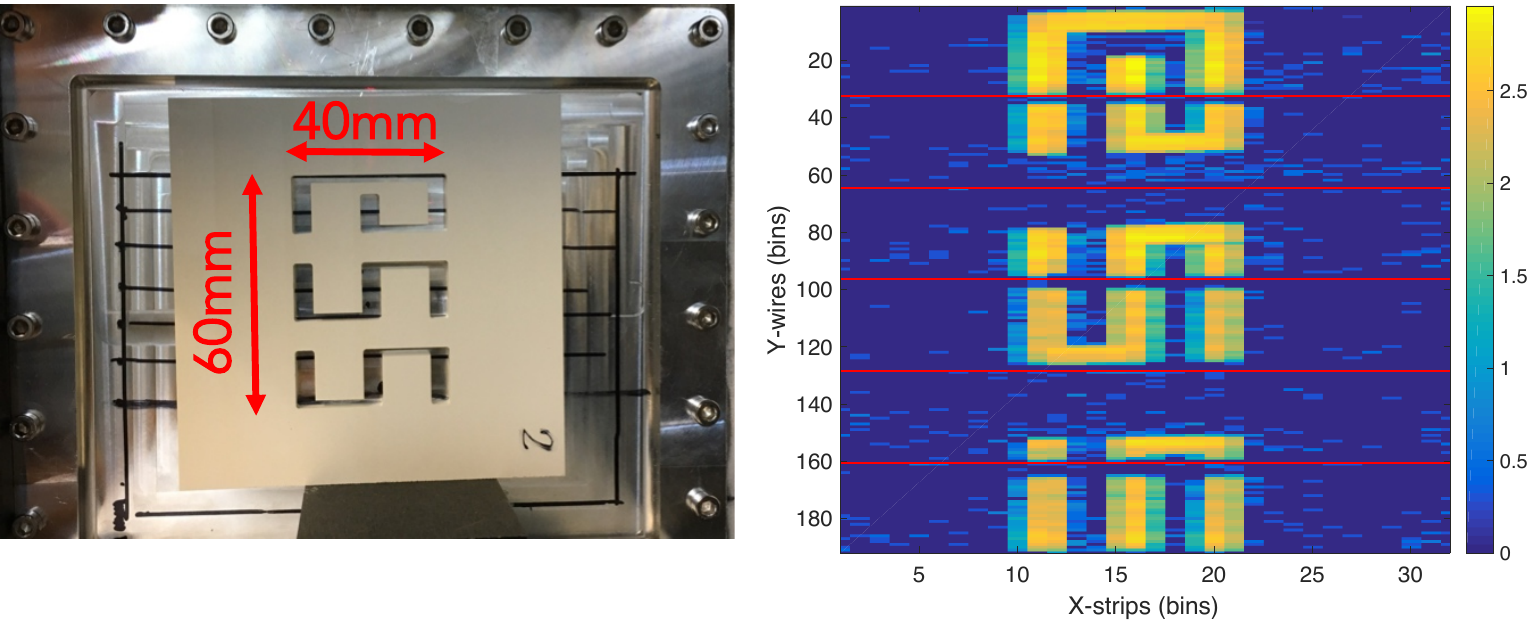}
\caption{\label{fig12} \footnotesize BN (Boron-Nitride) mask of the ESS logo positioned on the entrance window of the Multi-Blade detector (left). Raw reconstructed image of the ESS mask with the MAX algorithm showing all the channels (right). The color bar is shown in logarithmic scale and represents counts. The bin size on the X-axis is $4\,mm$ and on the Y-axis is $0.35\,mm$.}
\end{figure}

\begin{figure}[htbp]
\centering
\includegraphics[width=.49\textwidth,keepaspectratio]{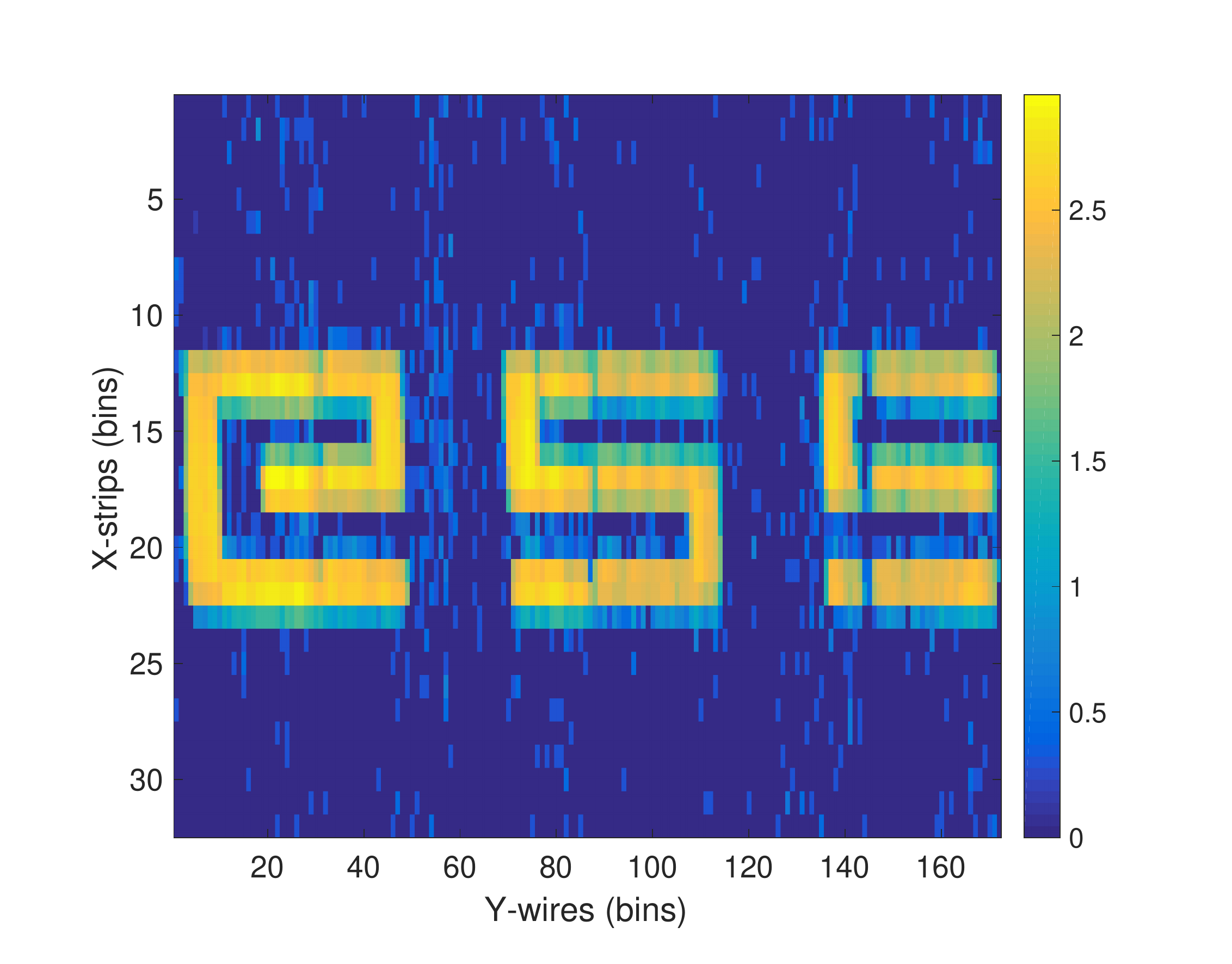}
\includegraphics[width=.49\textwidth,keepaspectratio]{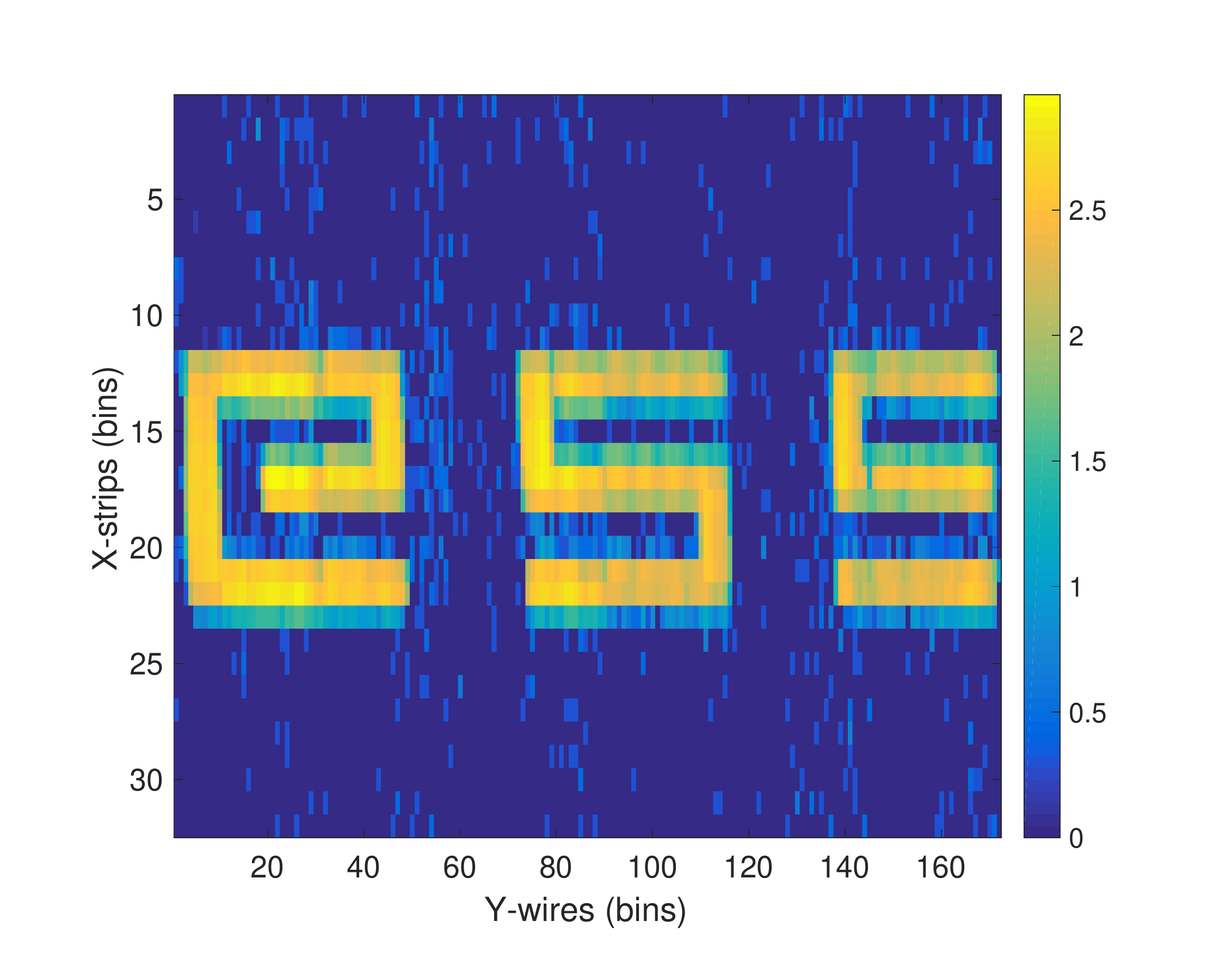}
\caption{\label{fig12bis} \footnotesize Raw reconstructed image of the ESS mask rotated by 90 degrees with shadowed channels removed. The correction is applied considering the $4\,m$ (left) and $2.3\,m$ (right) cassette array radius arrangement. The color bar is shown in logarithmic scale and represents counts. The bin size on the X-axis is $4\,mm$ and on the Y-axis is $0.35\,mm$.}
\end{figure}

\begin{figure}[htbp]
\centering
\includegraphics[width=.95\textwidth,keepaspectratio]{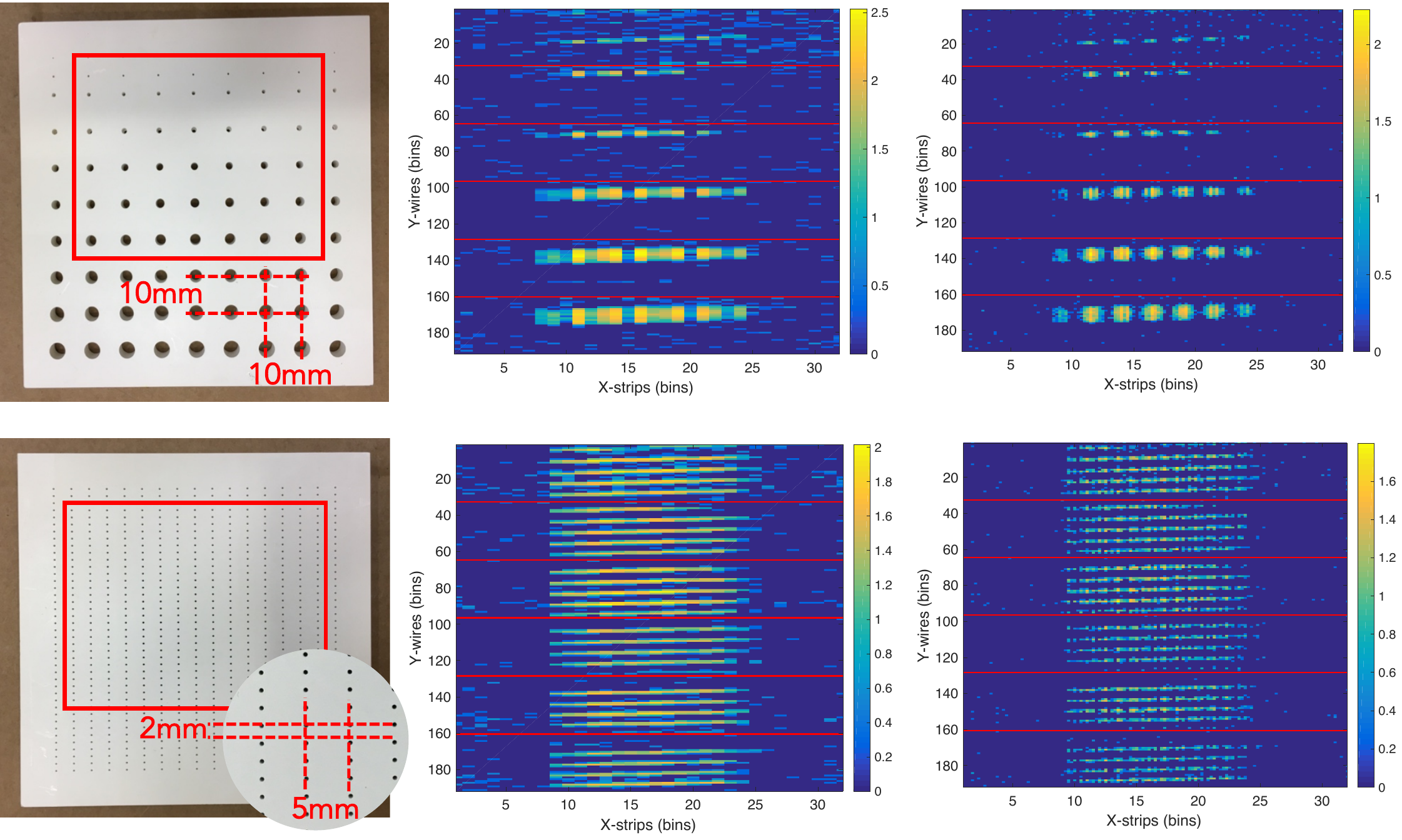}
\caption{\label{fig13} \footnotesize A picture of the BN masks (left) along with the reconstructed images with the MAX (center) and the CoG (right) algorithms. The color bar is shown in logarithmic scale and represents counts.}
\end{figure}

All the images shown in this section are gated in ToF above $8\, ms$ (2.5\AA). As the direct beam at the instrument is narrow and not enough to illuminate entirely the masks, a super-mirror (Fe/Si multi-layer) at the sample position was used. During the acquisition the sample angle was changed steadily in order to scan across the mask positioned at the detector window. 
\\ The first mask used in the tests is shown in Figure~\ref{fig12} (left) and it reproduces the ESS logo. This mask was used to study what happens at the overlap between cassettes. The image reconstructed with the MAX algorithm is also shown in Figure~\ref{fig12} (right). 
\\  In the present detector the cassettes are arranged over a circle of $4\,m$ radius, as for the ESTIA configuration, and each blade is positioned with $0.14$ degrees angle with respect to the adjacent unit. In this configuration the shadowing of each cassette on the neighbour is of about 3 wires, i.e. 3 bins (of 32). The bin size on the X-axis (strips) is $4\,mm$ and on the Y-axis (wires) is $0.35\,mm$: $\sin(5^o) \times 4\,mm$. The very last 3 wires of each cassette should not receive any neutron because they are physically hidden behind the neighbour blade (bins from 33 to 35, 65 to 67, 97 to 99, 129 to 131 and 161 to 163 in the image in Figure~\ref{fig12}). 
\\ On CRISP the Multi-Blade is positioned at $2.3\,m$ from the sample, and this distance was the maximum distance allowed by the CRISP setup. In this configuration, the sample-detector distance and the cassette array radius are not matching, and this does not allow to align all the cassettes of the detector at 5 degrees with respect to the incoming beam, but only one: the third from the top, bins from 64 to 96. Therefore, we expect the shadowing not to be constant across the cassettes. From the image in Figure~\ref{fig12} (right), the shadowing is between 2 and 3 wires (or bins from 33 to 35) for the top cassette and almost 5 wires for the bottom cassette (bins from 161 to 165). It is important to note that this is not a dead area of the detector and this shadowing effect can be removed from the images without losing any information. Figure~\ref{fig12bis} shows the reconstructed images (rotated) when this effect is removed. In the left image the shadows are removed according to the $4\,m$ array radius, i.e. 3 bins for each cassette. The larger the distance from the aligned cassette (bins from 64 to 96), the wider is the shadow. On the other hand, in the right image, the correction is applied according to the geometry at $2.3\,m$ on CRISP, thus the number of removed bins is not constant across the detector and the shadows disappear completely.
\\ As described in section~\ref{linea}, the blades overlap with a region of reduced sensitivity due to the reduced charge collection at the very first wire of each cassette (see the pink curve in the left plot in Figure~\ref{fig11}). When removing the shadows the reduced sensitivity area is visible in Figure~\ref{fig12bis} at the bin position 32, 89 and 143.
\\ Two more BN masks were used in the tests in order to compare the reconstruction algorithms. Figure~\ref{fig13} shows a picture of the masks (left) along with the reconstructed images with the MAX (center) and the CoG (right) algorithms. The red square marked on the picture of the masks represents the  area illuminated with neutrons that is reconstructed in the images. The CoG algorithm is applied only to reconstruct the strip position (X-axis in the plots) because the improvement on the wires is not significant. Note that the rows of holes in both the reconstructed images are slightly tilted because the detector was rotated by approximately 2 degrees with respect to the horizontal. 
\\ The reconstructed images in Figure~\ref{fig13} show a good separation across the wires (Y axis) being the resolution of the detector much better in this direction with respect to the other direction. With the MAX algorithm the separation across the strips (X axis) is barely visible. When the CoG algorithm is applied, the good separation between the $10\,mm$ holes of the top mask appears and the separation between the $5\,mm$ holes of the mask at the bottom start to be visible in the image. 
\\ Note that this is not a measurement of the spatial resolution of the detector which was fully characterized in~\cite{MIO_MB2017} because the actual footprint of the neutron spot on the detector is not defined due to the unknown divergence of the neutron beam through each hole of the mask. For instance, the actual $0.5\,mm$ diameter of the holes of the bottom mask is instead wider and unknown at the detector. 
\\ A figure of merit ($fom$) of the improvement of using the CoG algorithm with respect to the MAX algorithm can be defined as the average ratio between the counts at the peaks and the counts between the peaks. For the top image in Figure~\ref{fig13} we obtain a $fom\approx3.6$ for the MAX algorithm and $fom\approx5.7$ for the CoG algorithm when this is applied to the strip signals. 
\section{Conclusions and outlook}
The Multi-Blade is a promising alternative to $\mathrm{^3He}$ detectors for neutron reflectometry instruments in general. Neutron reflectometers are limited by the current detector technology mainly in terms of spatial resolution and counting rate capability. Moreover, the requirements for the detector technology set for the ESS reflectometers (ESTIA~\cite{INSTR_ESTIA,INSTR_ESTIA1,INSTR_ESTIA2} and FREIA~\cite{INSTR_FREIA,INSTR_FREIA2}) represent a clear challenge. Therefore, not only the ESS reflectometers, but also other reflectometers at other facilities in the world, can profit from the Multi-Blade detector technology.
\\ A Multi-Blade detector has been built and it has been tested on the CRISP~\cite{CRISP1} reflectometer at ISIS (Sciente \& Technology Facilities Council in UK~\cite{ISIS}). The aim of this test was to get a full technology demonstration in a reflectometry environment. Some characterization measurements on the technical aspects of the detector have been carried out. 
\\ The amount of material that can cause spurious scattering, and thus misaddressed events in the detector, has been reduced between the demonstrator shown in~\cite{MIO_MB2014} and the new versions~\cite{MIO_MB2017} and the one shown in this manuscript. By design, the $\mathrm{^{10}B_4C}$-layer in the current detector serves as a shielding, thus the substrate cannot be reached by neutrons that cannot be subsequently scattered within the detector. A spurious scattering effect was observed in the measurements and it was attributed to those low wavelength neutrons that are not absorbed by the $\mathrm{^{10}B_4C}$-coating, and being scattered by the substrate, they are detected in other cassettes. This effect vanishes if neutrons above 4\AA\, are selected. The actual coating thickness in the detector was $4.4\,\mu m$ rather than the recommended thickness of $7.5\,\mu m$. At the shortest wavelengths ($\approx 1$\AA) the $4.4\,\mu m$ coating is only $50\,\%$ efficient at absorbing neutrons. 
\\ The shortest wavelength that will be used at the ESS reflectometers is 2.5\AA\,(Table~\ref{tab1}), the nominal coating ($7.5\, \mu m$) at 2.5\AA\, is expected to be efficient ($>98\,\%$) at reducing the scattering as the $4.4\,\mu m$ coating used in these tests at 4\AA.
\\ Both the spatial and time dynamic range have been measured and the actual dynamic range of the instrument was reproduced. The spatial dynamic range between pixels is about four orders of magnitude (peak to tail) and the time dynamic range between subsequent time bins is approximately 3 orders of magnitude. This was limited by the dynamic range of the instrument where the test was performed~\cite{INSTR_OSMOND_CRISP}.
The measured detection efficiency of the Multi-Blade detector is in good agreement with the previous results~\cite{MIO_MB2017} and with the theoretical model~\cite{MIO_analyt}. It is approximately $45\,\%$ at the shortest wavelength (2.5\AA) that will be used at the ESS reflectometers.
\\ Stability measurement of the Multi-Blade over two weeks have been carried out at the Source Testing Facility (STF)~\cite{SF2,SF1} at the Lund University in Sweden. The counting rate in the detector is stable within $\pm 1.7\,\%$ during several days with a flow
that replace approximately two detector volumes ($\approx 60\,l$) per day. A clear correlation with the atmospheric pressure was found. In order to further improve the detector stability in time, the gas gain or thresholds would need to be adjusted according to the atmospheric pressure and temperature variations. An active feedback on the signal thresholds or on the high voltage as shown in~\cite{DET_Stability} or an off-line post-processing of the data can be considered~\cite{DMSC1,DMSC2}. 
\\ The overall variation of the gain, i.e. the uniformity, in the scanned cassette in both directions (wires and strips) is $10\,\%$. It shows improvements with respect to the previous detector described in~\cite{MIO_MB2017} because of the new substrates that replace the Aluminium and the individual readout which allows to operate the detector at lower gas gain with respect to charge division. This variation is correctable and therefore meets the instrument requirements.
\\ The Center-of-Gravity (CoG) algorithm is highly beneficial to improve the spatial resolution on the strips because it takes advantage from the higher multiplicity of the events with respect to the wires. 
\\ The overall test on the CRISP reflectometer demonstrates that the detector is matching the requirements to perform neutron reflectometry measurements. The Multi-Blade technology has been characterized at the instrument with a set of scientific measurements as well: the reflectivity of standard samples have also been measured and these results will be presented in a separate manuscript. 

\acknowledgments This work is being supported by the BrightnESS project, Work Package (WP) 4.2 (EU Horizon 2020, INFRADEV-3-2015, 676548) and carried out as a part of the collaboration between the European Spallation Source (ESS - Sweden), the Lund University (LU - Sweden), the Link\"{o}ping University (LiU - Sweden), the Wigner Research Centre for Physics (Hungary) and the University of Perugia (Italy).
\\ The work was supported by the Momentum Programme of the Hungarian Academy of Sciences under grant no. LP2013-60.
\\ The work was carried out in part at the Source Testing Facility, Lund University (LU - Sweden).
\\ The work originally started in the context of the collaboration between the Institut Laue-Langevin (ILL - France), the Link\"{o}ping University (LiU - Sweden) and the European Spallation Source (ESS - Sweden) within the context of the International Collaboration on the development of Neutron Detectors (www.icnd.org).
\\ The authors would like to thank the ISIS detector group for the support during the tests. 
\\ The authors thank the CRISP instrument scientists R. Dalgliesh and C. Kinane for providing the beam time and the instrument support necessary for this detector test.

\bibliographystyle{ieeetr}
\bibliography{BIBLIODB}
\end{document}